\documentclass[10pt,journal,compsoc]{IEEEtran}
\ifCLASSOPTIONcompsoc
\usepackage[nocompress,noadjust]{cite}
\else
\usepackage{cite}
\fi

\ifCLASSINFOpdf
\else
\fi

\usepackage[cmex10]{amsmath}
\usepackage{amssymb}  

\ifCLASSOPTIONcompsoc
  \usepackage[caption=false,font=footnotesize,labelfont=sf,textfont=sf]{subfig}
\else
  \usepackage[caption=false,font=footnotesize]{subfig}
\fi
\usepackage{fixltx2e}

\usepackage{stfloats}
\usepackage{url}

\usepackage{mathtools}

\newcommand{\norm}[1]{\lVert #1 \rVert}

\newcommand{\abs}[1]{\lvert #1 \rvert}

\DeclareMathOperator{\Var}{Var}
\DeclareMathOperator{\sgn}{sgn}
\DeclareMathOperator{\diag}{diag}

\newtheorem{theorem}{Theorem}

\newtheorem{proposition}[theorem]{Proposition}
\newtheorem{lemma}[theorem]{Lemma}

\newtheorem{definition}[theorem]{Definition}

\newtheorem{example}[theorem]{Example}

\newtheorem{remark}[theorem]{Remark}

\numberwithin{theorem}{section}

\usepackage{arydshln}
\usepackage{color}
\usepackage{accents}
\newcommand{\ubar}[1]{\underaccent{\bar}{#1}}
\usepackage{changepage}
\usepackage[all]{xy}

\begin{document}

%
\title{Stability of Spreading Processes over Time-Varying Large-Scale Networks%
}

\author{Masaki~Ogura,~\IEEEmembership{Member,~IEEE,}
    and~Victor~M.~Preciado,~\IEEEmembership{Member,~IEEE}
\IEEEcompsocitemizethanks{\IEEEcompsocthanksitem The authors are with the Department of Electrical and Systems Engineering at the University of Pennsylvania, Philadelphia, PA 19104. 
}
}

\IEEEtitleabstractindextext{%
\begin{abstract}
In this paper, we analyze the dynamics of spreading processes taking place over time-varying networks. A common approach to model time-varying networks is via Markovian random graph processes. This modeling approach presents the following limitation: Markovian random graphs can only replicate switching patterns with exponential inter-switching times, while in real applications these times are usually far from exponential. In this paper, we introduce a flexible and tractable extended family of processes able to replicate, with arbitrary accuracy, any distribution of inter-switching times. We then study the stability of spreading processes in this extended family. We first show that a direct analysis based on It\^o's formula provides stability conditions in terms of the eigenvalues of a matrix whose size grows exponentially with the number of edges. To overcome this limitation, we derive alternative stability conditions involving the eigenvalues of a matrix whose size grows linearly with the number of nodes. Based on our results, we also show that heuristics based on aggregated static networks approximate the epidemic threshold more accurately as the number of nodes grows, or the temporal volatility of the random graph process is reduced. Finally, we illustrate our findings via numerical simulations.
\end{abstract}

}

\maketitle

\IEEEdisplaynontitleabstractindextext

\IEEEpeerreviewmaketitle

\IEEEraisesectionheading{\section{Introduction}}
\IEEEPARstart{U}{nderstanding} spreading processes in complex networks is a
central question in the field of Network Science with applications in rumor
propagation~\cite{Lerman2010}, malware spreading~\cite{Garetto2003},
epidemiology, and public health~\cite{Tizzoni2012}. In this direction,
important advances in the analysis of spreading processes over
\emph{time-invariant} (TI) networks have been made during the last decade
(see~\cite{Pastor-Satorras2015a} and \cite{Nowzari2015a} for recent
surveys). Under the time-invariance assumption, we find several recent
breakthroughs in the literature, such as the rigorous analysis of
mean-field approximations \cite{Ahn2014}, the connection between the
eigenvalues of a network and epidemic thresholds
\cite{Ganesh2005,Chakrabarti2008,VanMieghem2009a}, new modeling tools for
analysis of multilayer networks \cite{DarabiSahneh2013}, and the use of
control and optimization tools to contain epidemic outbreaks
\cite{Wan2008IET,Preciado2013,Preciado2014, Drakopoulos2014}.

In practice, most spreading processes of practical interest take place in
networks with \emph{time-varying} (TV) topologies, such as human contact
networks, online social networks, biological, and ecological networks
\cite{Holme2012}. A naive, but common, approach to analyze dynamic
processes in TV networks is to build an aggregated static topology based on
time averages. When the time scales of the network evolution are comparable
to those of the spreading process, the static aggregated graph is not
well-suited to study the dynamics of TV networks, as pointed out in
\cite{Fefferman2007,Masuda2013}. Using extensive numerical simulations, it
has been observed that the speed of spreading of a disease in a TV network
can be substantially slower than in its aggregated static representation 
\cite{Vazquez2007}. This observation is also supported by the study of
spreading processes in a real TV network constructed from a mobile phone
dataset \cite{Karsai2011}. More recent studies point out the key role
played by the addition and removal of links \cite{Holme2014}, as well as
the distribution of contact durations between nodes~\cite{Vestergaard2014} in the dynamics of the spread. The works mentioned
above provide empirical evidence about the nontrivial effect that the
dynamics of the network has on the behavior of spreading processes.

Apart from these empirical studies, we also find several works providing theoretical support to these numerical observations. The authors in \cite{Volz2009,Schwarzkopf2010} derived the value of the epidemic threshold in particular types of TV networks, assuming that all nodes in the network present a homogeneous infection and recovery rates. In \cite{Perra2012a}, Perra et al. proposed a model of temporal network, called the activity-driven model, able to replicate burstiness in node/edge activity and analyzed contagion processes taking place in this model in \cite{Perra2012}.  Karsai et al. proposed in \cite{Karsai2014} a time-varying network using a reinforcement process able to replicate the emergence of strong and weak ties observed in a mobile call dataset. A wide and flexible class of TV network model, called \emph{edge-Markovian graphs}, was proposed in \cite{Clementi2008} and analyzed in \cite{Kiss2012}. In this model, edges appear and disappear independently of each other according to Markov processes. Edge-Markovian graphs have been used to model, for example, intermittently-connected mobile networks~\cite{Whitbeck2011,Maggi2014} and information spread thereon~\cite{Baumann2009}. Taylor et al.~derived in \cite{Taylor2012} the value of the epidemic threshold in edge-Markovian graphs and proposed control strategies to contain an epidemic outbreak, assuming homogeneous spreading and recovery rates.

It is worth remarking that, although most existing analyses of spreading processes over dynamic networks rely on the assumption that nodes present homogeneous rates and the Markov processes used to generate edge-switching signals are identical, these assumptions are not satisfied in real-world networks. {Furthermore, \label{discussion:nonexponential} most theoretical analyses mentioned above assume that edges appear and disappear in the network according to exponential distributions. However, as was found in several experimental studies~\cite{Cattuto2010,Stehle2011,Stehle2011a} utilizing radio-frequency identification devices (RFID), the probability distribution of contact durations in human proximity networks is far from exponential.} Therefore, it is of practical interest to develop tools to analyze spreading processes in TV networks with heterogenous rates and non-identical edge-switching signals able to replicate non-exponential switching patterns.\label{whyNon-Markovian} 

In this paper, we study spreading processes over TV networks with heterogeneous rates and non-identical edge-switching signals. First, we propose a wide class of TV networks where edges appear and disappear following \emph{aggregated Markov processes}~\cite{Fredkin1986}. Since the class of aggregated Markov processes contains all Markov processes, the proposed class includes all edge-Markovian graphs~\cite{Clementi2008}. {We furthermore observe that, due to the modeling flexibility of aggregated Markov processes, the proposed class of TV networks can replicate human-contact durations following non-exponential distributions with an arbitrary accuracy.} Second, we theoretically analyze the dynamics of spreading process over the proposed class of dynamic graphs. We model the dynamics of spreading processes on this class using a set of stochastic differential equations, which is a time-varying version of the popular $N$-intertwined SIS model~\cite{VanMieghem2009a}. Third, we derive conditions for these stochastic differential equations to be globally exponentially stable around the infection-free equilibrium, i.e., the infection `dies~out' exponentially fast over time. One of the main challenges in this analysis is to derive computationally tractable stability conditions. For example, as will be illustrated later, if we simply apply a recently proposed stability condition~\cite{Rami2014} based on the direct application of It\^o's formula for jump processes (see, e.g., \cite{Brockett2009}), we obtain a stability condition that requires the computation of the largest eigenvalue of a matrix whose size grows exponentially with the number of edges in the network. Hence, this condition is hard (if not impossible) to verify for large-scale networks. In this paper, we use spectral graph theory \cite{Chung2003a} to derive stability conditions for the set of stochastic difference equations in terms of the largest eigenvalues of a matrix whose size grows \emph{linearly} with the number
of nodes in the network.

This paper is organized as follows. In Section~\ref{sec:math}, we introduce the
notation and preliminary facts used in the paper and, then, introduce a flexible model of TV networks model called aggregated-Markovian random graph process.
In Section~\ref{sec:analysis}, we state computationally tractable stability
conditions for spreading processes taking place in this dynamic network model.
In Section~\ref{sec:example}, we illustrate our results with some numerical
examples. The proofs of the theorems are presented in Section~\ref{sec:pf}.

\section{Preliminaries and Problem Statement}\label{sec:math}

{This section starts by introducing notation in
Subsection~\ref{sec:notation}. Then, in Subsection~\ref{sec:SwitchModel}, we
review preliminary facts about spreading processes over static networks.
In Subsection~\ref{sec:model}, we introduce the class of TV network models under consideration. We finalize this section by stating the problem to be studied in
Subsection~\ref{subsec:problem}.}

\subsection{Notation} \label{sec:notation}

For a positive integer $n$, define the set $[n] = \{1, \dotsc, n\}$. For $c\in
\mathbb{R}$, define $c^- = (\abs{c} - c)/2$. Let $\sgn(\cdot)$ denote the sign
function over $\mathbb{R}$, which can be extended to matrices by entry-wise
application. The Euclidean norm of $x\in \mathbb{R}^n$ is denoted by $\norm{x}$.
Let $I_n$ denote the $n\times n$ identity matrix. A square matrix is said to be
\emph{Metzler} if its off-diagonal entries are nonnegative. The \emph{spectral
abscissa} of a square matrix $A$, denoted by $\eta(A)$, is defined as the
maximum real part of its eigenvalues. We say that $A$ is \emph{Hurwitz stable}
if $\eta(A) < 0$. Also, we define the \emph{matrix measure} \cite{Desoer1972} of
$A$ by $\mu(A) = \eta(A+A^\top)/2$. For a square random matrix $X$, its
expectation is denoted by $E[X]$ and its variance is defined by $\Var (X) =
E[(X-E[X])^2]$. The diagonal matrix with diagonal elements $a_1$, $\dotsc$,
$a_n$ is denoted by $\diag(a_1, \dotsc, a_n)$.

A directed graph is defined as the pair $\mathcal{G}=(\mathcal{V},\mathcal{E})$,
where $\mathcal{V}=\{ v_{1},\dots,v_{n}\} $ is a set of nodes\footnote{For
simplicity in notation, we sometimes refer to node $v_i$ as $i$.} and $\mathcal{E}\subseteq\mathcal{V}\times\mathcal{V}$ is a set of
directed edges (also called \emph{diedges} or \emph{arcs}), defined as ordered
pairs of nodes. By convention, we say that $(v_{j},v_{i})$ is a directed edge
from $v_{j}$ pointing towards~$v_{i}$. The \emph{adjacency matrix} $A=[a_{ij}]$
of a directed graph $\mathcal{G}$ is defined as the $n\times n$ matrix such that
$a_{ij}=1$ if $(v_{j},v_{i}) \in \mathcal E$, and $a_{ij}=0$ otherwise.

A stochastic process $\sigma$ taking values in a set $\Gamma$ is said to be an \emph{aggregated Markov process} (see, e.g., \cite{Fredkin1986}) if there exists a time-homogeneous Markov process $\theta$ taking values in a set~$\Lambda$ and a function $f\colon \Lambda \to \Gamma$ such that $\sigma = f(\theta)$. Throughout the paper, it is assumed that $\Lambda$ is finite, $f$ is surjective, and all the Markov processes are time-homogeneous. We say that a Markov process $\theta$ is \emph{irreducible} if all the states in $\Lambda$ can be reached from any other state in $\Lambda$ (see, e.g., \cite{Cinlar1975} for more details). We say that an aggregated Markov process~\mbox{$\sigma = f(\theta)$} is \emph{irreducible} if $\theta$ is irreducible. We will use the next lemma in our proofs.

\begin{lemma}[\cite{Cinlar1975}]\label{lem:markov}
Let $\theta$ be an irreducible Markov process taking values in a finite set
$\Lambda$. Then, $\theta$ has a unique stationary distribution $\pi$. Moreover,
\begin{equation*}
\pi(\lambda) = \lim_{t\to\infty}\Pr(\theta(t) = \lambda),
\end{equation*}
for every $\lambda \in \Lambda$ and $\theta(0)\in \Lambda$.
\end{lemma} 

\subsection{Spreading Process over Static Networks}\label{sec:SwitchModel}

In this subsection, we review a model of spreading processes over static networks called the \emph{Heterogeneous Networked SIS} (HeNeSIS) model, which is an extension of the popular $N$-intertwined SIS model~\cite{VanMieghem2009a} to the case of nodes with heterogeneous rates. This model can be described using a continuous-time Markov process, as follows. Let $\mathcal G$ be a directed graph, where nodes in $\mathcal G$ represent individuals and diedges represent interactions between them (which we consider to be directed). At a given time~$t \geq 0$, each node can be in one of two possible states: {\it susceptible} or {\it infected}. We define the variable $X_i(t)$ as $X_i(t)=1$ if node~$i$ is infected at time $t$, and $X_i(t)=0$ if $i$ is susceptible. In the HeNeSIS model, when a node~$i$ is infected, it can randomly transition to the susceptible state according to a Poisson process with rate~$\delta_i > 0$, called the {\it recovery rate} of node~$i$. On the other hand, if node $i$ is in the susceptible state, it can randomly transition to the infected state according to a Poisson process with rate~$\beta_i \sum_{i=1}^n a_{ij} X_j(t)$, where $\beta_i > 0$ is called the {\it infection rate} of node $i$. In other words, the rate of transition of node $i$ from susceptible to infected is proportional to the number of its infected neighbors.

Since the above Markov process has a total of $2^n$ possible states (two states 
per node), its analysis is very hard for arbitrary contact networks of large 
size. A popular approach to simplify the analysis of this type of Markov 
process is to consider {upper-bounding linear models.} Let $A$ 
denote the adjacency matrix of $\mathcal G$, and define $p = (p_1, \dotsc, 
p_n)^\top$, $P = \diag(p_1, \dotsc, p_n)$, $B = \diag(\beta_1, \dotsc, 
\beta_n)$, and $D = \diag(\delta_1, \dotsc, \delta_n)$. Then, 
{it is known that} the solutions $p_i(t)$ ($i=1$, $\dotsc$, 
$n$) 
of 
the linear differential equation
\begin{equation}
\dot p = (B A-D)p
\label{LinearMFA}
\end{equation}
upper-bounds the evolution of $E[X_i(t)]$ ($i=1$, $\dotsc$, $n$) from the exact Markov process with $2^n$ states \cite{VanMieghem2009a}. Thus, if the dynamics in \eqref{LinearMFA} is globally exponentially stable, the infection dies out exponentially fast in the exact Markov process \cite{Preciado2014,Ahn2014}. In the case of homogeneous infection and recovery rates, i.e., $\beta_i=\beta$ and $\delta_i=\delta$ for all $i$, the dynamics~\eqref{LinearMFA} is globally exponentially stable if and only if \cite{VanMieghem2009a}
\begin{equation}
\frac{\beta}{\delta} < \frac{1}{\eta(A)}.
\label{LinearMFA:TI}
\end{equation}

Apart from the continuous-time Markov process described above, we can also model the dynamics of a spread using a discrete-time networked Markov chain with an exponential number of states \cite{Chakrabarti2008}. As in the continuous-time case, we can show \cite{Chakrabarti2008} that the solution of the linear difference equation
\begin{equation}\label{eq:def:disc:N:pre:linear}
p(k+1) = (BA + I-D)p(k),
\end{equation}
where $p = (p_1, \dotsc, p_n)^\top$, upper-bounds the probabilities 
$\Pr(\text{$i$ is infected at time $k$})$ ($i=1$, $\dotsc$, $n$). As in the 
continuous-time case, it turns
out that if the discrete-time linear system in \eqref{eq:def:disc:N:pre:linear}
is globally exponentially stable, the infection dies out exponentially fast in
the exact Markov process \cite{Ahn2014}. In the case of homogeneous infection
and recovery rates, the system in \eqref{eq:def:disc:N:pre:linear} is stable if
and only if \eqref{LinearMFA:TI} holds~\cite{Chakrabarti2008}.

\subsection{Aggregated-Markovian Random Graph Processes} \label{sec:model}

In this subsection, we introduce two new models of TV networks: the
\emph{aggregated-Markovian edge-independent} (AMEI) and the
\emph{aggregated-Markovian arc-independent} (AMAI) models. As we mentioned
above, these models generalize the class of edge-Markovian time-varying networks
\cite{Clementi2008}. Let $\mathcal G = \{\mathcal G(t)\}_{t\geq 0}$ be a
stochastic graph process taking values in the set of directed graphs with $n$
nodes. Let $A(t)$ denote the adjacency matrix of $\mathcal G(t)$ for each
$t\geq 0$. The class of dynamic random graphs studied in this paper is defined
as follows.

\begin{definition}\label{def:am:edge}
Consider a collection of aggregated Markov processes $\sigma_{ij} =
g_{ij}(\theta_{ij})$ where $g_{ij}$ are $\{0, 1\}$-valued functions and
$\theta_{ij}$ are stochastically independent Markov processes for $1\leq i<j\leq
n$.  An \emph{aggregated-Markovian edge-independent} (AMEI) network is a random
graph process in which $A_{ij} = A_{ji} = \sigma_{ij}$ for $i<j$ and $A_{ii} =
0$ for all $i\in [n]$. If $\sigma_{ij}$ is irreducible for all possible
pairs~$(i, j)$, then we say that $\mathcal G$ is \emph{irreducible}.
\end{definition}

\begin{figure}
\newcommand{\myrule}{\rule[-.27cm]{0cm}{.7cm}}%
\centering
\centerline{\xymatrix@!C{
 *+=[Fo]{\myrule c_1} \ar[r]^-{\text{\normalsize$p_1$}} \ar@/^.75pc/@(r,lu)[ddrrr]^-{\text{\normalsize$q_1$}}& *+=[Fo]{\myrule c_2} \ar[r]^-{\text{\normalsize$p_2$}} \ar@(r,u)[ddrr]^-{\text{\normalsize$q_2$}}& \cdots \ar[r]^-{\text{\normalsize$p_{n-1}$}} & *+=[Fo]{\myrule c_n} \ar@(rd,ur)[dd]^-{\text{\normalsize$q_n$}}
\\
 & &  &
\\
*+=[Fo]{\myrule d_m} \ar@(ul,dl)[uu]^-{\text{\normalsize$s_m$}}&\cdots\ar[l]^-{\text{\normalsize$r_{m-1}$}}   & *+=[Fo]{\myrule d_2} \ar[l]^-{\text{\normalsize$r_2$}} \ar@(l,d)[lluu]^-{\text{\normalsize$s_2$}} &*+=[Fo]{\myrule d_1} \ar[l]^-{\text{\normalsize$r_1$}} \ar@/^.75pc/[uulll]^{\text{\normalsize$s_1$}}
 }}
\caption{{A Markov process described in Example \ref{ex:modelingAbility}.}}
\label{fig:example:aggregatedMarkov}
\end{figure}
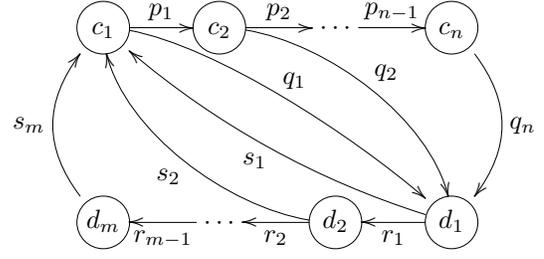

In other words, the upper-triangular entries of the time-varying adjacency
matrix $A(t)$ of an AMEI graph are independent aggregated Markov processes. The
lower triangular entries are constructed via symmetry and the diagonal entries
are zero. Hence, the graph is undirected by construction and does not contain
self-loops. AMEI graphs extend the class of  edge-Markovian graphs
\cite{Clementi2008} and allow us to model a wider class of edge processes. For
example, in an edge-Markovian graph, the time it takes for an edge to switch
from connected to disconnected (or vice versa) follows an exponential
distribution. In contrast, in an AMEI graph, we can design the function $g_{ij}$
and the Markov process $\theta_{ij}$ to fit any desired distribution for the
contact durations with arbitrary precision, as illustrated in the following example:

\begin{example}\label{ex:modelingAbility}
{Let $\theta$ be the Markov process over the state space~$S = \{c_1, \dotsc, c_n, d_1, \dotsc, d_m\}$ presenting the state-transition diagram shown in Fig.~\ref{fig:example:aggregatedMarkov}. Consider an edge modeled by the aggregated Markov process $\sigma = f(\theta)$, where the function $f\colon S \to \{0, 1\}$ is defined by $f(c_i) = 1$ and $f(d_j) = 0$ for all $i\in [n]$ and $j\in[m]$. Therefore, the edge is active if and only if $\theta$ is in one of the states $c_1$, $\dotsc$, $c_n$. From the diagram in Fig.~\ref{fig:example:aggregatedMarkov}, we see that the edge is active from the time $\theta$ enters $c_1$ until the time $\theta$ arrives to $d_1$. It can be proved that the time elapsed between these two events follows a Coxian probability distribution, which forms a dense subset in the set of nonnegative random variables~\cite{Schassberger1970}. Therefore, by appropriately choosing the transition rates~$p_1$, $\dotsc$, $p_{n-1}$, $q_1$, $\dotsc$, $q_n$, it is possible to tune the distribution followed by the time the edge is active in order to follow any distribution with an arbitrary accuracy (when $n$ is large enough). From the symmetry of the diagram in Fig.~\ref{fig:example:aggregatedMarkov}, we can also tune the parameters $r_1$, $\dotsc$, $r_{m-1}$, $s_1$, $\dotsc$, $s_m$ so that the time the edge is inactive follows any given probability distribution with an arbitrary accuracy.}
\end{example}

In the following definition, we introduce a directed version of AMEI graphs.

\begin{definition}\label{def:am:arc}
Consider a collection of aggregated Markov processes $\sigma_{ij} =
g_{ij}(\theta_{ij})$ where $g_{ij}$ are $\{0, 1\}$-valued functions and
$\theta_{ij}$ are stochastically independent Markov processes for $i, j\in [n]$
and $i\neq j$.  An \emph{aggregated-Markovian arc-independent} (AMAI) network is
a random graph process in which $A_{ij} = \sigma_{ij}$ for $i\neq j$ and $A_{ii}
= 0$ for all $i\in [n]$. If $\sigma_{ij}$ is irreducible for all possible
pairs~$(i, j)$, then we say that $\mathcal G$ is \emph{irreducible}.
\end{definition}

Once the class of aggregated-Markovian networks is introduced, we analyze
spreading processes taking on these random graph processes.

\subsection{Problem Statement}\label{subsec:problem}

In this paper, we address the following question: \emph{Under what conditions does a spreading process taking place in a TV network with aggregated-Markovian link dynamics die out exponentially fast?} In other words, we consider the problem of analyzing the stability of the upper-bounding linear model \eqref{LinearMFA} when the graph is an aggregated-Markovian network. In this case, the entries of the adjacency matrix are random processes and, therefore, \eqref{LinearMFA} is the following stochastic linear differential equation:
\begin{equation*}
\Sigma : \dot p(t) = \left(B A(t) -D\right)p(t), 
\end{equation*}
where $A(t)$ is a random matrix process. In the following section, we will
derive conditions under which the disease-free equilibrium of $\Sigma$ is
stable, i.e., the infection probabilities converge to zero as $t\to\infty$ 
{exponentially fast}, in
the following sense:

\begin{definition}\label{def:stability}
Consider a random graph process $\mathcal{G}$ defined by either an AMEI or AMAI
graph. We say that the disease-free equilibrium of $\Sigma$ is \emph{almost
surely exponentially stable} if {there exists $\lambda > 0$ such that
\begin{equation*} 
{\Pr \left( \limsup_{t\to\infty} \frac{\log\norm{p(t)}}{t} \leq -\lambda  \right) = 1, }
\end{equation*}
for all initial states $p(0) = p_0$ and $\sigma_{ij}(0) = \sigma_{ij, 0}$. The supremum of $\lambda$ satisfying the above condition is called the \emph{decay rate}.}
\end{definition}

\section{Stability of Spreading Processes in Random Graph Processes}
 \label{sec:analysis}

In this section, we state the main results in this paper. In particular, we
derive conditions for the spreading model $\Sigma$ to be stable when the network
structure varies according to an AMEI (or AMAI) random graph process. In
Subsection~\ref{subsec:analysis_exp}, we apply a stability
condition~\cite{Rami2014} based on the direct application of the It\^o formula
for jump processes to derive stability conditions in terms of the largest
eigenvalue of a matrix whose size depends exponentially on the number of edges
in the graph. Due to the exponential size of this matrix, this condition is hard
(if not impossible) to apply in the analysis of large-scale networks. Motivated
by this limitation, we propose in Subsection \ref{subsec:analysis_lin} an
alternative approach using tools from the theory of random matrices. In
particular, we derive stability conditions in terms of the largest eigenvalue of
a matrix whose size depends linearly on the number of nodes. In
Subsection~\ref{subsec:disc}, we extend our results from continuous-time Markov
processes to spreading processes modeled as discrete-time Markov chains.

\subsection{Stability Conditions: Exponential Matrix Size} \label{subsec:analysis_exp}

In this subsection, we show that the result in \cite[Theorem~7.1]{Rami2014}
based on the direct application of the It\^o formula for jump processes results
in stability conditions that are not well-suited for large-scale graphs. We
illustrate this idea through the analysis of the stability of the spread when
$\mathcal G$ is an AMAI random graph process. For simplicity in our exposition,
we here temporarily assume that all the processes $\sigma_{ij}$ are Markovian;
i.e., the mappings $g_{ij}$ are the identities.

To state our claims, we need to introduce the following notations. Let $\mathcal
G$ be an AMAI random graph process (Definition \ref{def:am:arc}). Let $\mathcal
E$ be the set of (time-varying) directed edges in $\mathcal G$ (i.e., those
directed edges such that $\sigma_{ij}$ is not the zero stochastic process). Let
$m = \abs{\mathcal E}$ be the number of diedges in~$\mathcal G$ and label the
diedges using integers $1,\dotsc,m$. Recall that these edges are not always
present in the graph, but they appear (link is `on') and disappear (link is
`off') according to a collection of aggregated-Markov processes. Therefore, at a
particular time instant, only a subset of the edges in $\mathcal E$ are present
in the graph. The set of present edges can be represented by an element of $\{0,
1\}^m$, since we have $m$ possible edges that can be either `on' of `off'. In
other words, at a particular time, the contact network is one of the $2^m$
possible subgraphs of $\mathcal G$.

To analyze the resulting dynamics, we consider the set of $2^m$ possible
subgraphs of $\mathcal G$ and label them using integers $\{1,\ldots,2^m\}$. We
define the one-to-one function $\chi \colon \{1,\ldots,2^m\} \to \{0, 1\}^m$, as
the function that maps a particular subgraph of $\mathcal G$ into the binary
string in $\{0, 1\}^m$ that indicates the subset of present edges in the
subgraph. For the subgraph labeled $\ell$, we denote by $\chi_\ell$ the
corresponding binary sequence and by $\chi_\ell (k)$ the $k$-th entry in this
sequence. In other words, $\chi_\ell (k)=1$ means that the edge labeled $k$ in
$\mathcal G$ is present in the subgraph labeled $\ell$. Furthermore, we denote
by $F_\ell \in \{0,1\}^{n\times n}$ the adjacency matrix representing the
structure of the subgraph labeled $\ell$.

Given two labels $\ell, \ell'\in [2^m]$, define the edit distance~$d(\ell,
\ell')$ to be the minimum number of links that must be added and/or removed from
the subgraph labeled $\ell$ to construct the subgraph labeled $\ell'$, i.e.,
\mbox{$d(\ell, \ell')=|\{k \in [m] \colon \chi_\ell(k) \neq
\chi_{\ell'}(k)\}|$}. If $d(\ell, \ell')=1$ (i.e., the subgraphs differ in a
single diedge), we define $k_{\ell, \ell'}$ as the (only) index $k$ such that
$\chi_\ell(k) \neq \chi_{\ell'}(k)$ (i.e., the label of the single edge that
must added/removed). Finally, assume that the transition probabilities of the
Markov process $\sigma_{ij}$ ($(i, j) \in \mathcal E$) are given by
\begin{equation*}
\begin{aligned}
\Pr(\sigma_{ij}(t+h) = 1 \mid \sigma_{ij}(t) = 0) &= u_{ij}h + o(h), 
\\
\Pr(\sigma_{ij}(t+h) = 0 \mid \sigma_{ij}(t) = 1) &= v_{ij}h + o(h),
\end{aligned}
\end{equation*}
for some nonnegative constants $u_{ij}$ and $v_{ij}$. We define $u(k) =
u_{i_kj_k}$ and $v(k) = v_{i_kj_k}$ for each $k\in [m]$. 

Under the above notations, the following stability condition readily follows from \cite[Theorem~7.1]{Rami2014}: 

\begin{proposition}\label{prop:exp_size_stab}
Let $\mathcal G$ be an AMAI random graph process. Define $\Pi \in
\mathbb{R}^{2^m \times 2^m}$ entry-wise, as follows. For $\ell\neq\ell'$,
\begin{equation*}
\Pi_{\ell\ell'} = \begin{cases}
u(k_{\ell,\ell'}),  & \text{if $d(\chi_\ell, \chi_{\ell'}) = 1$ and $\chi_\ell(k_{\ell,\ell'}) = 0$}, 
\\
v(k_{\ell,\ell'}),  & \text{if $d(\chi_\ell, \chi_{\ell'}) = 1$ and $\chi_\ell(k_{\ell,\ell'}) = 1$}, 
\\
0, & \text{otherwise.}
\end{cases}
\end{equation*}
For $\ell=\ell'$, we let $\Pi_{\ell\ell} = -\sum_{\ell'\neq \ell}
\Pi_{\ell\ell'}$. Then, for every $i$, the probability that $i$ is infected at
time $t$ in the HeNeSIS model converges to zero exponentially fast as
$t\to\infty$ if the $(n2^m)\times (n2^m)$ matrix
\begin{equation}\label{massive_matrix}
\Pi \otimes I_n + \bigoplus_{\ell=1}^{2^m} (BF_{\ell} - D), 
\end{equation}
where $\otimes$ denotes the Kronecker product of matrices, is Hurwitz stable.
\end{proposition}

As stated before, this stability condition is not applicable to large-scale
networks, since it requires the computation of the eigenvalue with the largest
real part of a matrix whose size increases exponentially with the number of
edges in the graph $\mathcal G$. In the following subsection, we present
alternative stability conditions that are better-suited for stability analysis
of large networks.

\subsection{Stability Analysis: Linear Matrix Size} \label{subsec:analysis_lin}

In this subsection, we present sufficient conditions for almost sure {exponential}
stability of the disease-free equilibrium of $\Sigma$ when the underlying
time-varying network is described by an AMEI (or AMAI) random graph
process. For clarity in our exposition, we state and discuss our main
results in this section, leaving the details of the proofs for Subsections~\ref{app:arc} and \ref{app:edge}.

We now state our main results. Let us consider two positive constants $b$ and
$d$, and define the decreasing function $\kappa_{b,d} \colon [0, \infty) \to (0,
n]$ by 
\begin{equation}\label{kappa_function}
\kappa_{b,d}(s) = n e^{s/b} \left(
\frac{b s+d}{d}
\right)^{-\frac{b s+d}{b^2}}, 
\end{equation}
where $n$ is the number of the nodes in the network. The following theorem
provides conditions for a spreading process taking place in an AMAI random graph
process to be almost surely {exponentially} stable.

\begin{theorem}\label{thm:links}
Let $\mathcal G$ be an irreducible AMAI random graph process and define $\bar A
\in \mathbb{R}^{n\times n}$ component-wise, as follows
\begin{equation}\label{eq:def:barA}
\bar A_{ij} = 
\lim_{t\to\infty} \Pr(A_{ij}(t) = 1).
\end{equation}
Let $\bar \beta = \max_{1\leq i\leq n} \beta_i$, $\ubar \delta = \min_{1\leq 
i\leq n}\delta_i$, and
\begin{equation*}
\begin{aligned}
\Delta_1 &= \max_{1\leq i\leq n} \sum_{j=1}^n\left(\beta_i^2\bar A_{ij}(1-\bar A_{ij}) + \beta_j^2 \bar A_{ji}(1-\bar A_{ji}) \right),
\\
c_1 &= \mu(B(\sgn \bar A) - D) - \frac{\kappa_{\bar \beta, \Delta_1}^{-1}\!(1)}{2},
\\
\bar s_1 &= 2\ubar \delta + 2c_1^-.
\end{aligned}
\end{equation*}
Define
\begin{equation} \label{eq:tao_c}
\tau_A
=
\underset{ s\in ( \kappa_{\bar \beta, \Delta_1}^{-1}\!(1), \bar s_1 ] 
}{\text{maximize}} \left(-\frac{s + 2c_1 \kappa_{\bar \beta, \Delta_1}(s)}{2(1- 
\kappa_{\bar \beta, \Delta_1}(s))}\right).
\end{equation}
If
\begin{equation}\label{eq:threshold:arc}
\mu(B\bar A - D) < \tau_A,  
\end{equation}
then the disease-free equilibrium of $\Sigma$ is almost surely 
{exponentially}
stable with a decay rate greater than or equal to 
\begin{equation*}
-\mu(B\bar A - D)\left(1-\kappa_{\bar \beta, \Delta_1}(s^*)\right) - (s^*/2) - c_1 
\kappa_{\bar 
\beta, 
\Delta_1}(s^*), 
\end{equation*}
where $s^*$ denotes the optimal value of $s$ in the maximization 
problem~\eqref{eq:tao_c}.
\end{theorem}

\begin{IEEEproof}
See Subsection~\ref{app:arc}.
\end{IEEEproof}

\begin{remark}
The existence of the limit in \eqref{eq:def:barA} is guaranteed by the
irreducibility of $\mathcal G$ as will be discussed in Subsection~\ref{app:arc},
where we will also give an explicit representation of the limit.
\end{remark}

The major computational cost for checking the stability condition in
\eqref{eq:threshold:arc} comes from that of finding the matrix measure
$\mu(B\bar A-D)$. Since the matrix $B\bar A-D$ has size~$n$, we can compute
$\mu(B\bar A-D)$ in $O(nm^2)$ operations using Lanczos's
algorithm~\cite{Lanczos1950}, where $n$ and $m$ are the number of nodes and
diedges in $\mathcal G$. In contrast, computing the dominant eigenvalue of the
$n2^m$-dimensional matrix in~\eqref{massive_matrix} requires $O(n m^2  2^{m})$
operations, since this matrix has $O(m 2^m)$ nonzero entries. Therefore, the
result in Theorem~\ref{thm:links} (based on random graph-theoretical results)
represents a major computational improvement with respect to the result in
Proposition \ref{prop:exp_size_stab} (based on a direct application of the It\^o
formula for jump processes \cite{Rami2014}).

Above, we have analyzed the stability of spreading processes in
aggregated-Markovian arc-independent (AMAI) networks. In the rest of the
subsection, we derive stability conditions for spreading processes taking place
in aggregated-Markovian edge-independent (AMEI) networks. The next theorem is
the AMEI counterpart of Theorem~\ref{thm:links}.

\begin{theorem}\label{thm:edges}
Let $\mathcal G$ be an irreducible AMEI random graph process. Let
\begin{equation}\label{eq:Delta2}
\begin{aligned}
\Delta_2 &= \max_{1\leq i\leq n} \sum_{j=1}^n \left(\beta_i\beta_j \bar A_{ij}(1-\bar A_{ij}) \right),
\\
c_2 &= \eta( B (\sgn \bar A) - D) - \kappa_{\bar \beta, \Delta_2}^{-1}\!(1), 
\\
\bar s_2 &= \ubar{\delta} + c_2^-.
\end{aligned}
\end{equation}
Define
\begin{equation} \label{eq:tao_E}
\tau_E=\underset{ s \in (\kappa_{\bar \beta, \Delta_2}^{-1}\!(1), \bar s_2] 
}{\text{maximize}} \left(-\frac{s+c_2\kappa_{\bar \beta, 
\Delta_2}(s)}{1-\kappa_{\bar \beta, \Delta_2}(s)}\right).
\end{equation}
If 
\begin{equation}\label{eq:threshold:edges}
\eta(B\bar A - D) < \tau_E,
\end{equation}
then the disease-free
equilibrium of $\Sigma$ is almost surely {exponentially} stable 
{with a decay rate greater than or equal to
\begin{equation}\label{eq:decayRate:AMEI}
-\eta(B\bar A - D)\left(1-\kappa_{\bar
\beta, \Delta_2}(s^*)\right) - s^* - c_2\kappa_{\bar \beta, \Delta_2}(s^*), 
\end{equation}
where $s^*$ denotes the optimal value of $s$ in the maximization 
problem~\eqref{eq:tao_E}.} 
\end{theorem}

\begin{IEEEproof}
See Subsection~\ref{app:edge}. 
\end{IEEEproof}

Theorems \ref{thm:links} and \ref{thm:edges} provide threshold conditions for
stability of spreading processes in time-varying networks when agents present
heterogeneous infection and recovery rates, $\beta_i$ and $\delta_i$,
respectively. In the particular case of networks with homogeneous rates, i.e.,
$\beta_i=\beta$ and $\delta_i=\delta$ for all $i$, our results have an appealing
interpretation that can be related with existing results in the literature. The
following theorem states a threshold stability condition in the homogeneous
case:

\begin{theorem}\label{thm:edge:homo}
Let $\mathcal G$ be an irreducible AMEI random graph process. Assume that
$\beta_i=\beta >0$ and $\delta_i =\delta >0$ for all $i\in [n]$. Let
\begin{equation}
\begin{aligned}
\Delta_3 &= \max_{1\leq i\leq n} \sum_{j=1}^n \left( \bar A_{ij} (1-\bar A_{ij})\right), \label{Delta3}
\\
c_3 &= \eta(\sgn \bar A) - \kappa_{1, \Delta_3}^{-1}\!(1), 
\\
\bar s_3 &= \frac{\delta}{\beta} + c_3^-.
\end{aligned}
\end{equation}
Define
\begin{equation} \label{eq:tao_H}
\xi_H=\underset{ s \in (\kappa_{1, \Delta_3}^{-1}\!(1), \bar s_3] 
}{\text{maximize}} \frac{1 - (\beta/\delta)\left(s + c_3 \kappa_{1, 
\Delta_3}(s)\right)}{1-\kappa_{1, \Delta_3}(s)}.
\end{equation}
If
\begin{equation} \label{eq:Cond}
 \frac{\beta}{\delta} < \frac{\xi_H}{\eta(\bar A)}, 
\end{equation}
then the disease-free equilibrium of $\Sigma$ is almost surely 
{exponentially} stable {with decay rate greater than or equal to 
\begin{equation*}
(\delta/\beta) - \eta(\bar A) - s^* - (c_3-\lambda_3)\kappa_{1,
\Delta_3}(s^*), 
\end{equation*}
where $s^*$ denotes the optimal value of $s$ in the optimization 
problem~\eqref{eq:tao_H}.}
Moreover, if 
\begin{equation}\label{eq:beta/delta>eta(barA)}
\frac{\beta}{\delta} \geq \frac{1}{\eta(\sgn \bar A)}, 
\end{equation}
then $\xi_H < 1$.
\end{theorem}

\begin{IEEEproof}
See Subsection~\ref{app:edge}.
\end{IEEEproof}

\begin{remark}
Since $A(t) \leq \sgn \bar A$ entry-wise for every $t\geq 0$, the solution
of $\Sigma$ is always upper-bounded by that of the deterministic differential equation
\mbox{$\dot p(t) = (\beta \sgn \bar A  - \delta I) p(t)$}, 
whose solution converges to zero as $t\to\infty$ if and only if $\beta/\delta <
1/\eta(\sgn \bar A)$. Therefore, if the
inequality~\eqref{eq:beta/delta>eta(barA)} is false, then the disease-free
equilibrium of $\Sigma$ is almost surely {exponentially} stable and, hence, we do not need to
check condition~\eqref{eq:Cond}. In other words, condition
\eqref{eq:beta/delta>eta(barA)} guarantees that stability analysis of $\Sigma$
is not a trivial problem.
\end{remark}

As shown in Subsection~\ref{sec:SwitchModel}, the epidemic threshold in a static network of agents with homogeneous rates is given by $\beta/\delta < 1/\eta( A)$. Condition \eqref{eq:Cond} provides a similar epidemic threshold for time-varying networks in terms of the spectral abscissa of $\bar A$, which can be interpreted as an aggregated static network based on long-time averages. Furthermore, $\xi_H$ is a multiplicative factor that modifies the epidemic threshold corresponding to the aggregated static network. {Below, we derive explicit expressions of $\eta(\bar A)$ for a few particular examples of time-varying graph models found in the literature:}

\begin{figure}[!t]
\centering \includegraphics[width=3.5cm]{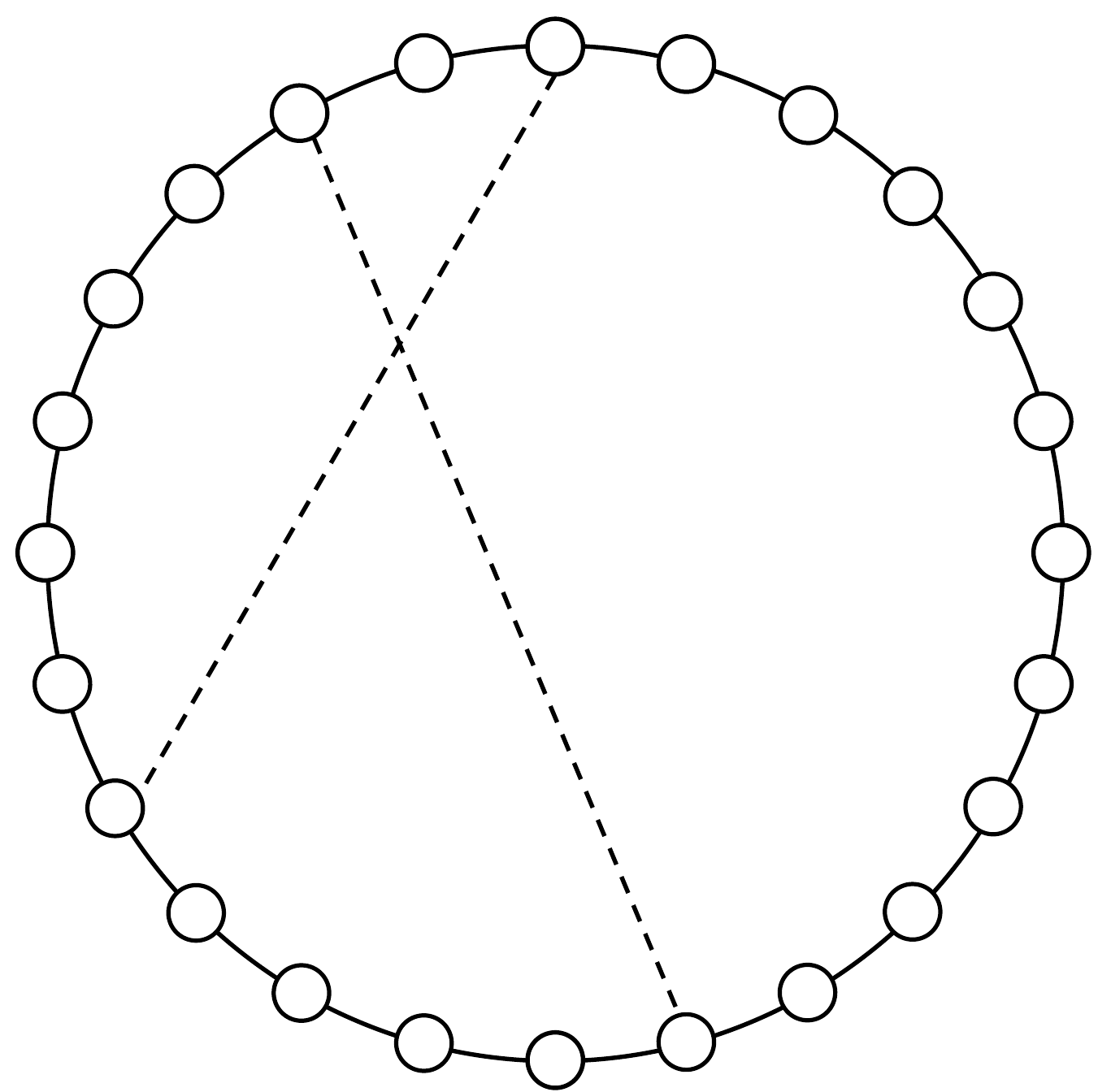} \caption{{Dynamic small-world network. Solid lines represent static edges, while dashed lines represent temporarily active edges modeled by aggregated-Markov processes.}}
\label{Fig.DynamicSW}
\end{figure}

\begin{example}[Dynamic small-world networks]\label{ex:SW}
{We consider a time-varying version of the small-world network model studied in~\cite{Saramaki2005}. The network consists of $n$ nodes and $n$ static edges in the set~$\mathcal E_0 = \{(1,2), (2,3), \dotsc, (n-1,n), (n,1)\}$ (see Fig. \ref{Fig.DynamicSW}). Apart from these static edges, we also consider a collection of dynamic edges that switch on and off over time. In particular, we allow any other pair of nodes $(i, j) \notin \mathcal E_0$ ($i\neq j$) to be dynamically connected according to an irreducible aggregated Markov process $a_{ij} = g_{ij}(\theta_{ij})$. For simplicity in our analysis, we assume that the stationary probability of an edge $(i, j)\notin \mathcal E_0$ being active, defined by $r_{ij} = \lim_{t\to\infty}\Pr(a_{ij}(t) = 1)$, equals a constant $r > 0$ independent of $i$ and $j$. In this case, the matrix~$\bar A$ takes the form
\begin{equation*}
\bar A_{ij} = 
\begin{cases}
0,  & \text{if $i=j$, }\\
1,  & \text{if $(i, j) \in \mathcal E_0$, }\\
r,  & \text{otherwise.}
\end{cases}
\end{equation*}
The spectral abscissa of~$\bar A$ is given by $1 + r(n-2)$. Therefore, the stability condition~\eqref{eq:Cond} reads $\beta/\delta < \xi_H/(1+r(n-2))$.}
\end{example}

\begin{example}[Edge-Markovian graph]\label{ex:EM}
{In this example, we consider the edge-Markovian graph model proposed in~\cite{Clementi2008}. In this model, all the undirected edges~$(i, j)$ ($i<j$) are time-varying and modeled by independent $\{0, 1\}$-valued Markov processes~$\theta_{ij}$ sharing the same activation rate~$q>0$ and de-activation rate~$r>0$. In this case, we have
\begin{equation*}
\bar A_{ij} = 
\begin{cases}
0,  & \text{if $i=j$, }\\
q/(q+r),  & \text{otherwise,}
\end{cases}
\end{equation*}
and therefore $\eta(\bar A) = (n-1)q/(q+r)$. The stability condition~\eqref{eq:Cond} hence reads $\beta/\delta < \xi_H (q+r)/((n-1)q)$.}
\end{example}

\begin{figure*}[!t]
\subfloat[\ \ \ \ \ ]
{\includegraphics[width=0.32\textwidth]{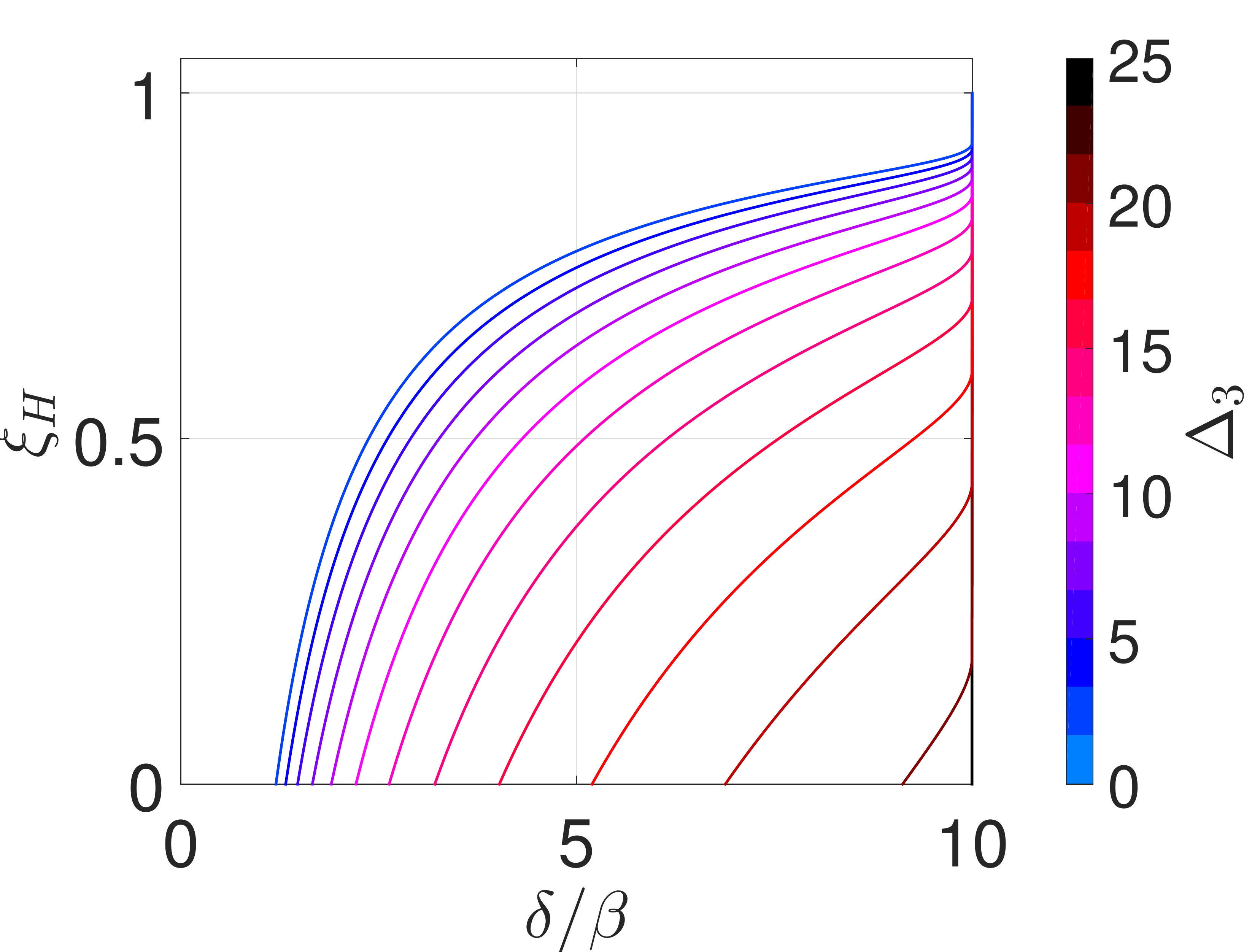}}
\hfill
\subfloat[\ \ \ \ \ \ \ ]
{\includegraphics[width=0.32\textwidth]{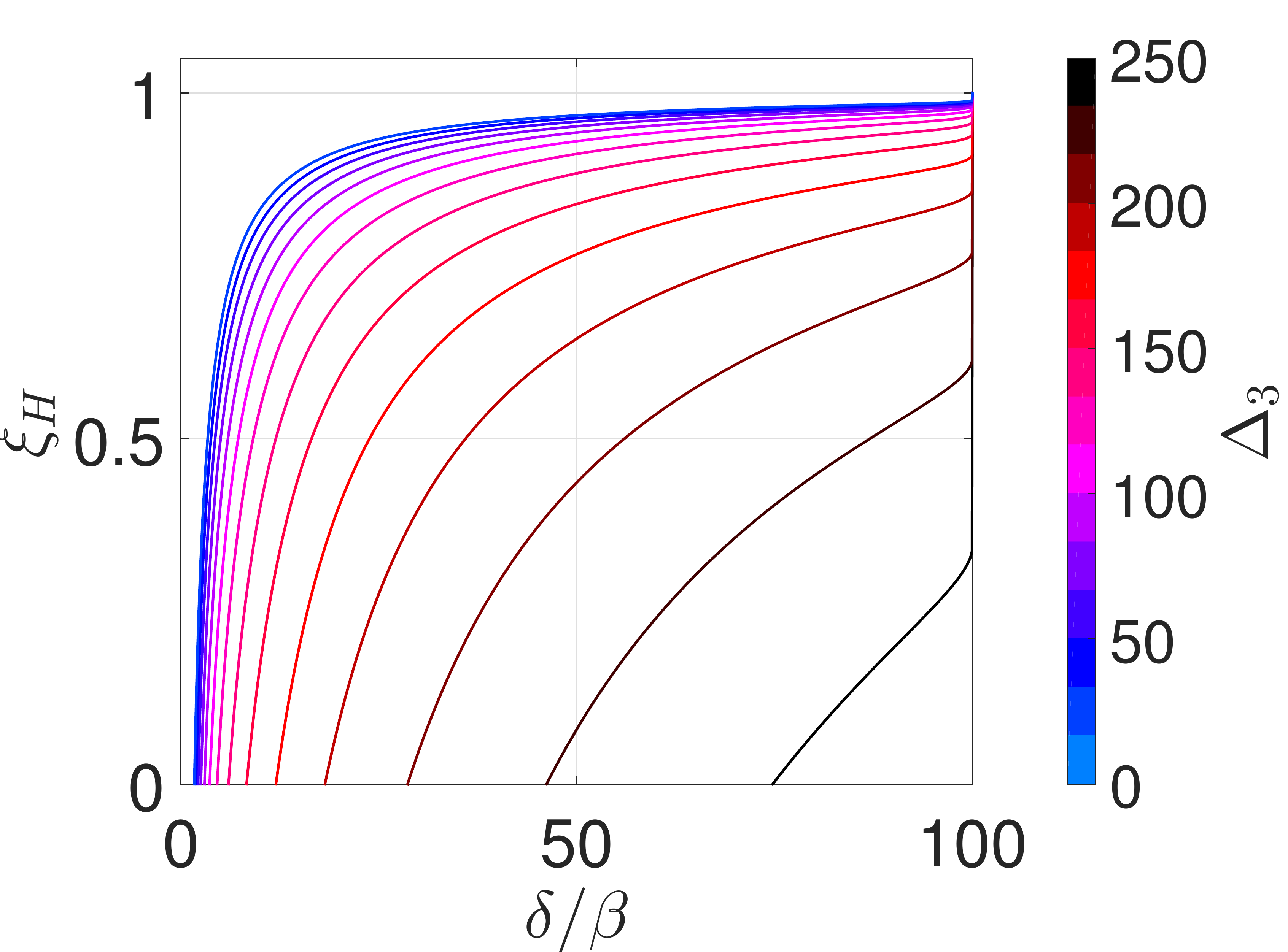}}
\hfill
\subfloat[\ \ \ \ \ \ \ \ \ ]
{\includegraphics[width=0.32\textwidth]{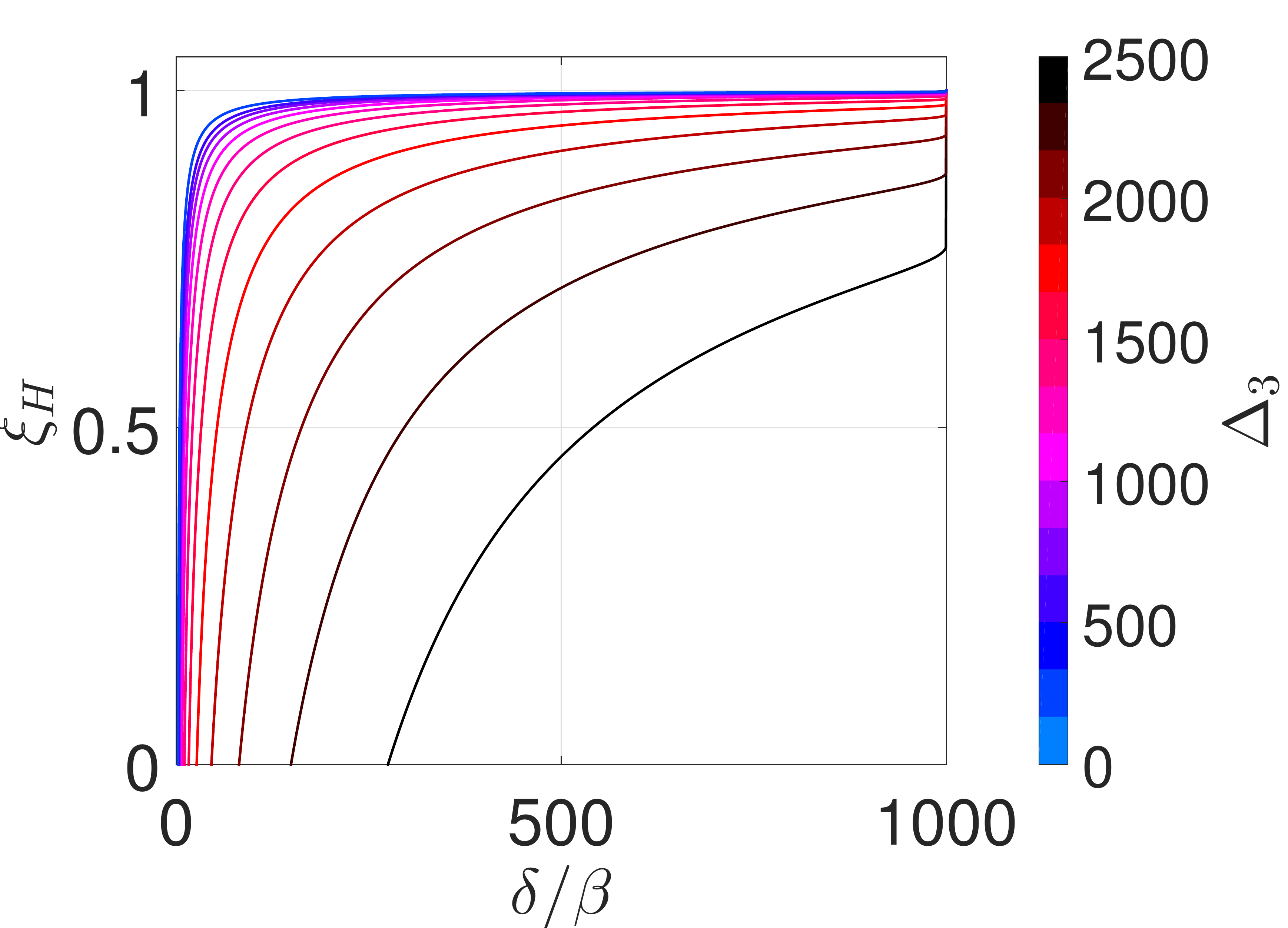}}
\caption{Plots of $\xi_H$ as $\delta/\beta$ and $\Delta_3$ vary for the following cases: ({a}) $n = 100$ and $\eta(\sgn \bar A) = 10$, ({b}) $n = 1,000$ and $\eta(\sgn \bar A) = 100$, and ({c}) $n = 10,000$ and $\eta(\sgn \bar A) = 1,000$.}
\label{fig:varyRatioAnd}
\end{figure*}

{In what follows, we elaborate on the behavior of the factor $\xi_H$}. In particular, we will analyze its dependence on $\Delta_3$ defined in \eqref{Delta3}. Notice that~$\Delta_3$ is zero when $\bar A_{ij}$ is either $0$ or~$1$ for all pairs~$(i,j)$, which corresponds to the case of a static, deterministic graph $\mathcal G$. In contrast, the maximum value of $\Delta_3$ is achieved when $\bar A_{ij}=1/2$, i.e., edges are either `on' or `off' with equal probability (asymptotically). Therefore, the term $\Delta_3$ can be interpreted as a measure of structural variability, in the long run. In Fig.~\ref{fig:varyRatioAnd}, we illustrate the behavior of the factor~$\xi_H$ as we modify the network size $n$, the epidemic ratio $\delta/\beta$, and the uncertainty measure $\Delta_3$. We observe that the smaller the value of $\Delta_3$, the larger the value of~$\xi_H$. In other words, as we reduce the structural variability (i.e., we reduce $\Delta_3$), we obtain a better approximation using an aggregated static model (i.e., $\xi_H$ approaches $1$). Furthermore, notice that the value of $\xi_H$ tends to $1$ as $n$ increases. This indicates that the aggregated static network approximates the epidemic threshold more accurately as the network grows.

In this section, we have introduced our main theoretical results, namely, we have studied the epidemic threshold in the wide class of aggregated-Markovian random graph processes. Before we illustrate our results with some numerical simulations in Section~\ref{sec:example}, we will discuss the case when the epidemics is modeled as a discrete-time stochastic process.

\subsection{Discrete-Time Epidemic Dynamics}\label{subsec:disc}

In this subsection, we briefly present a discrete-time version of the stability
analysis provided above. We define a discrete-time dynamic random graph as a
stochastic process $\mathcal G = \{\mathcal G(k)\}_{k=0}^\infty$ taking values
in the set of directed graphs with $n$ nodes. In the discrete-time case,
aggregated-Markovian edge- and arc-independent dynamic random graphs are defined
in the same way as in the continuous-time case. Let us consider the
discrete-time epidemic model in \eqref{eq:def:disc:N:pre:linear} taking place in
the dynamics random graph $\mathcal G$. The resulting dynamics is described in
the following stochastic difference equation:
\begin{equation*}
\Sigma_d : p(k+1) = (BA(k) + I-D)p(k), 
\end{equation*}
where $A(k)$ denotes the adjacency matrix of $\mathcal G(k)$.

We define almost sure {exponential} stability of the disease-free equilibrium of $\Sigma_d$
in the same way as we did for the continuous-time case:

\begin{definition}
Let $\mathcal G$ be a discrete-time edge-independent dynamic random graph. We say that the disease-free equilibrium of $\Sigma_d$ is \emph{almost surely {exponentially} stable} if {there exists $\lambda >  0$ such that
\begin{equation*}
\Pr\left( \limsup_{k\to\infty}\frac{\log\norm{p(k)}}{k} \leq - \lambda\right) = 1
\end{equation*}
for all $p(0) = p_0 \in [0, 1]^n$ and $\sigma_{ij}(0) = \sigma_{ij, 0}$ ($1\leq i<j\leq n$).
We call the supremum of $\lambda$ satisfying the above condition the \emph{decay rate}.}
\end{definition}

For simplicity in our exposition, we focus only on the edge-independent case
(AMEI graph process), since the analysis is identical for the arc-independent
case. The next theorem provides a sufficient condition for almost sure {exponential} stability
of the disease-free equilibrium of $\Sigma_d$.

\begin{theorem}\label{thm:disc:edge}
Consider an irreducible and aperiodic\footnote{We say that an AMEI random graph
process in discrete-time is {\it aperiodic} if all the aggregated Markov
processes in the adjacency matrix of the graph are the images of aperiodic
Markov chains.} AMEI random graph process  $\mathcal G$ in discrete-time. Let
$\lambda_4=\eta(B\bar A + I - D)$ and $M_{\max} = B(\sgn \bar A) + I - D$.
Define
\begin{equation}\label{eq:tau_D}
\tau_D
=
\underset{0\leq s\leq 1-\lambda_4}{\text{maximize}}
\left(
\left(\frac{\lambda_4}{\eta(M_{\max})} \right)^{\kappa_{\bar \beta,\Delta_2}(s)}- s
\right), 
\end{equation}
where $\Delta_2$ is defined in \eqref{eq:Delta2}. 
If
\begin{equation}\label{eq:disccond}
\lambda_4 < \tau_D, 
\end{equation}
then the disease-free equilibrium of $\Sigma_d$  is almost surely {exponentially} stable {with decay rate greater than or equal to}
\begin{equation}\label{eq:decayRateHomo}
\gamma_D 
= 
-\log(
\lambda_4 + s^*
)
- \kappa_{\bar \beta,\Delta_2}(s^*) \log \frac{\eta(M_{\max})}{\lambda_4},
\end{equation}
where $s^*$ denotes the value of $s$ giving the solution to the optimization
problem~\eqref{eq:tau_D}.
\end{theorem}

\begin{IEEEproof}
See Subsection~\ref{app:disc}. 
\end{IEEEproof}

\section{Numerical Simulations}\label{sec:example}

\begin{figure*}[!t]
\begin{minipage}{.49\textwidth}
\vspace{-.055cm}	
\includegraphics[width=\textwidth]{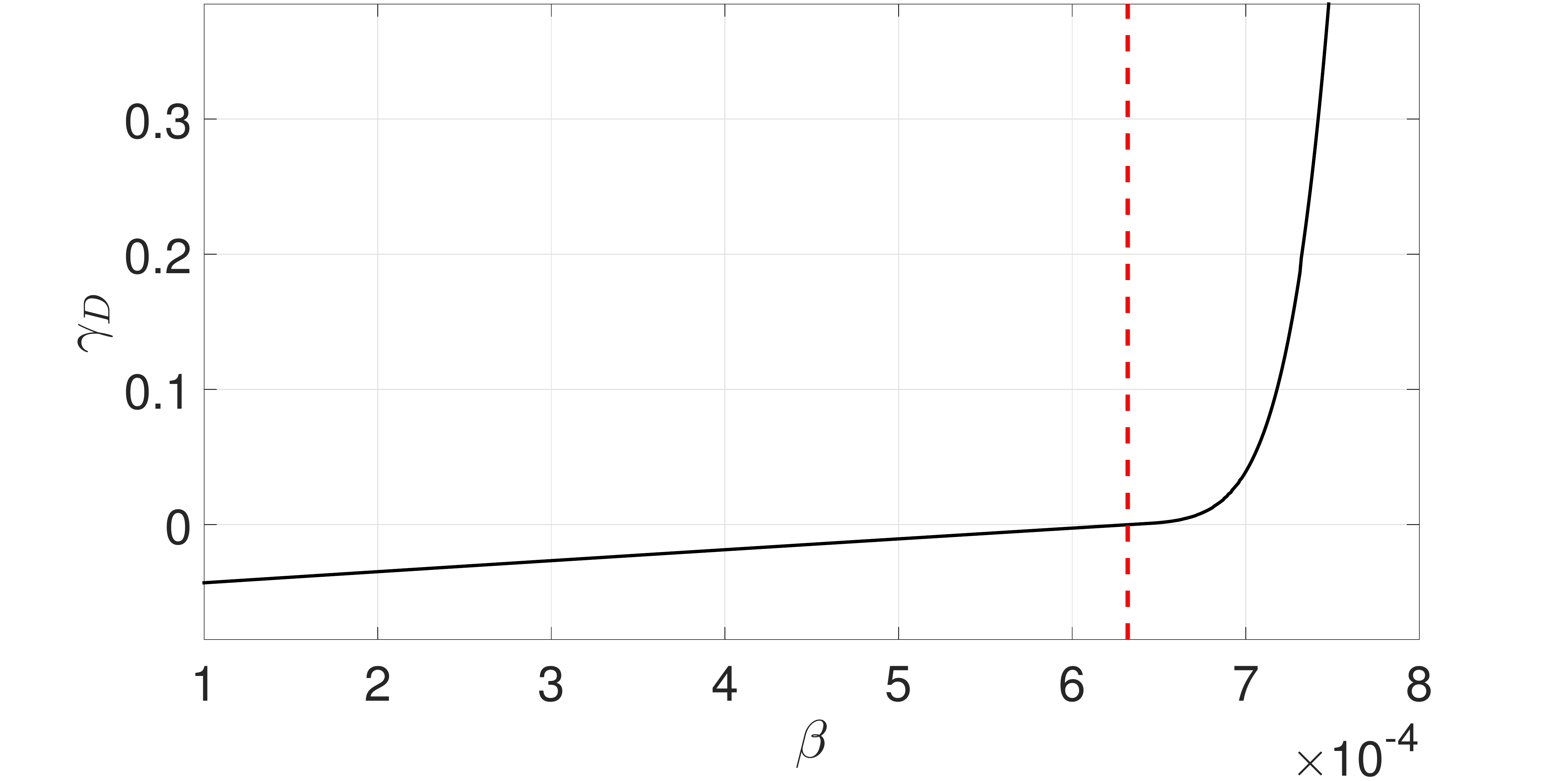} 
\caption{{The upper bound $\gamma_D$ of decay rates. Red dashed line shows the epidemic threshold given by Theorem~\ref{thm:disc:edge}.} \newline ~ \newline } \label{fig:upperBounds}
\end{minipage}\hfill
\begin{minipage}{.49\textwidth}
\includegraphics[width=\textwidth]{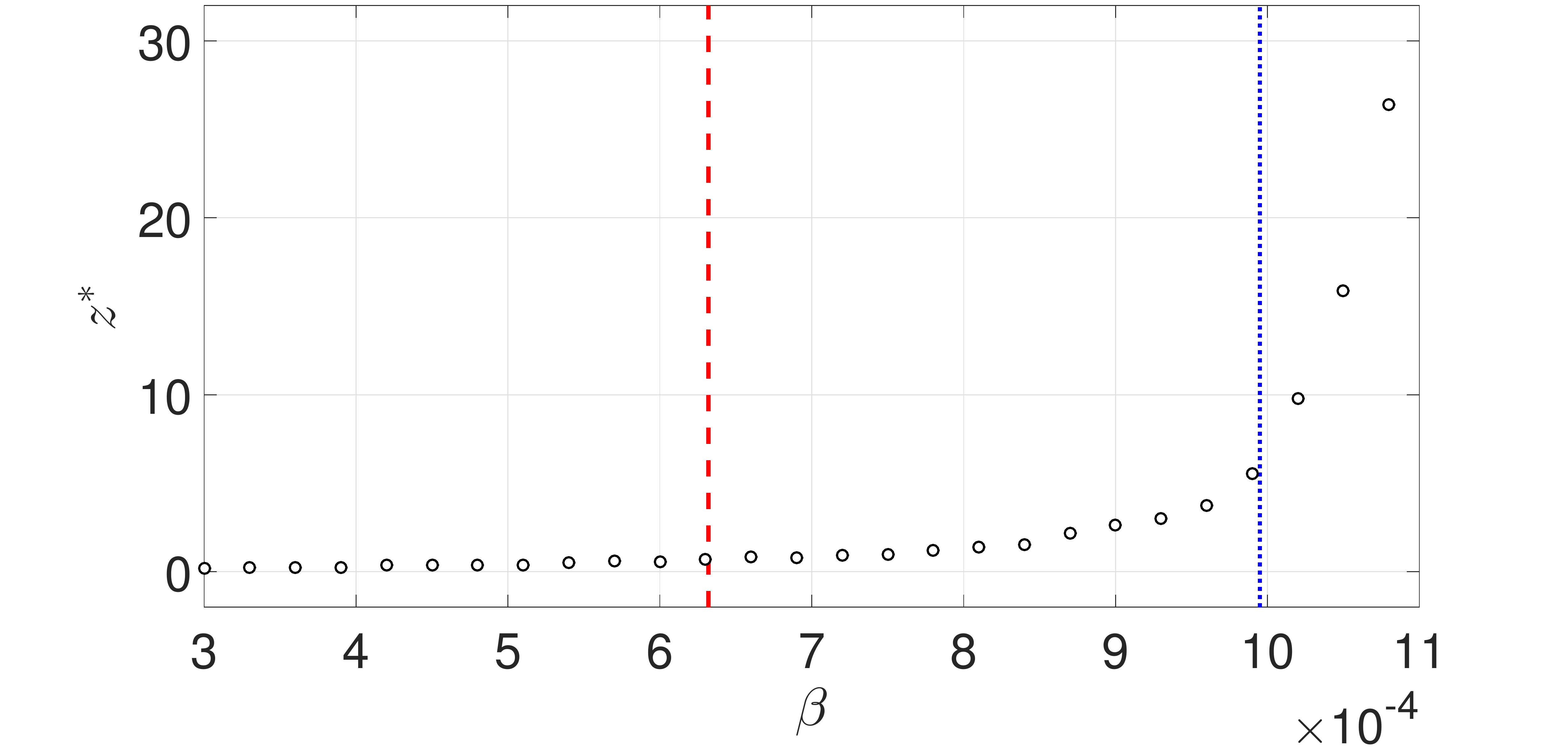}
\caption{The averaged numbers of the infected nodes at time $k = 1000$ versus $\beta$. Red dashed line:  Epidemic threshold predicted by Theorem~\ref{thm:disc:edge}. Blue dotted line:  Epidemic threshold predicted by using aggregated static network $\bar A$. }
\label{fig:beta-vs-fraction}
\end{minipage}
\\
\vspace{.3cm}
\subfloat[$\beta = 6.0\times 10^{-4}$]
{\includegraphics[width=0.3\textwidth]{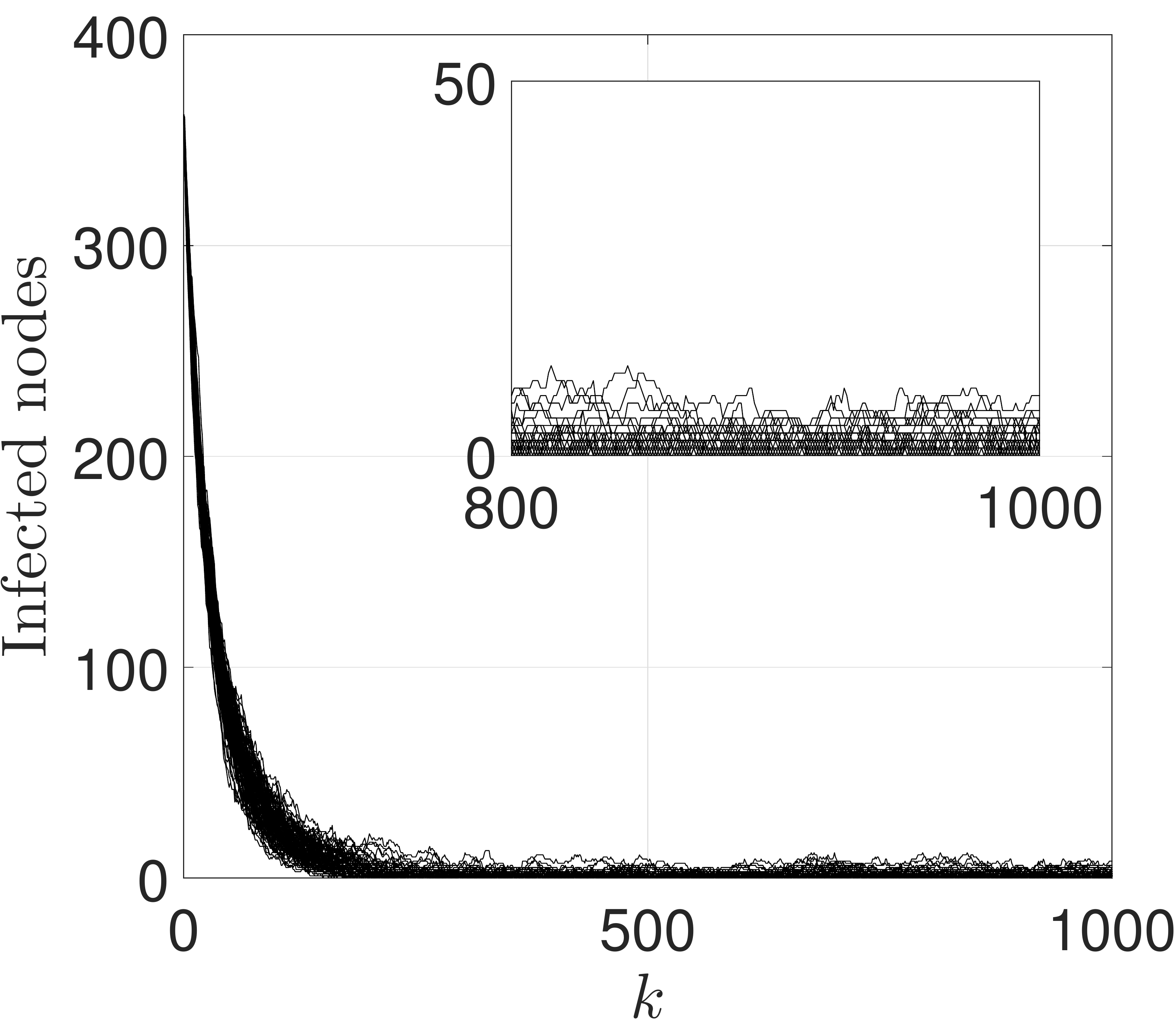}}
\hfill
\subfloat[$\beta = 7.5\times 10^{-4}$]
{\includegraphics[width=0.3\textwidth]{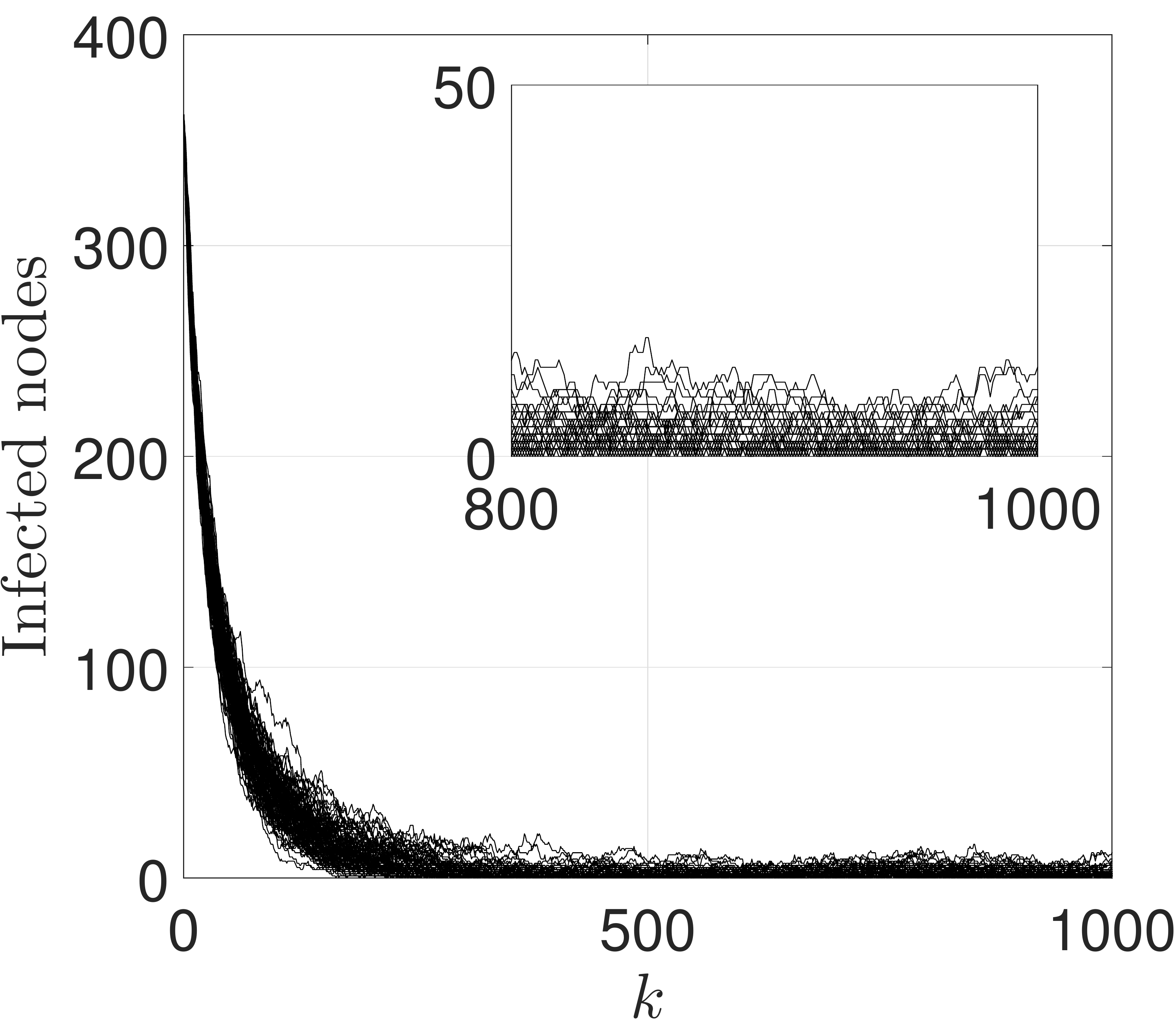}}
\hfill
\subfloat[$\beta = 9.0\times 10^{-4}$]
{\includegraphics[width=0.3\textwidth]{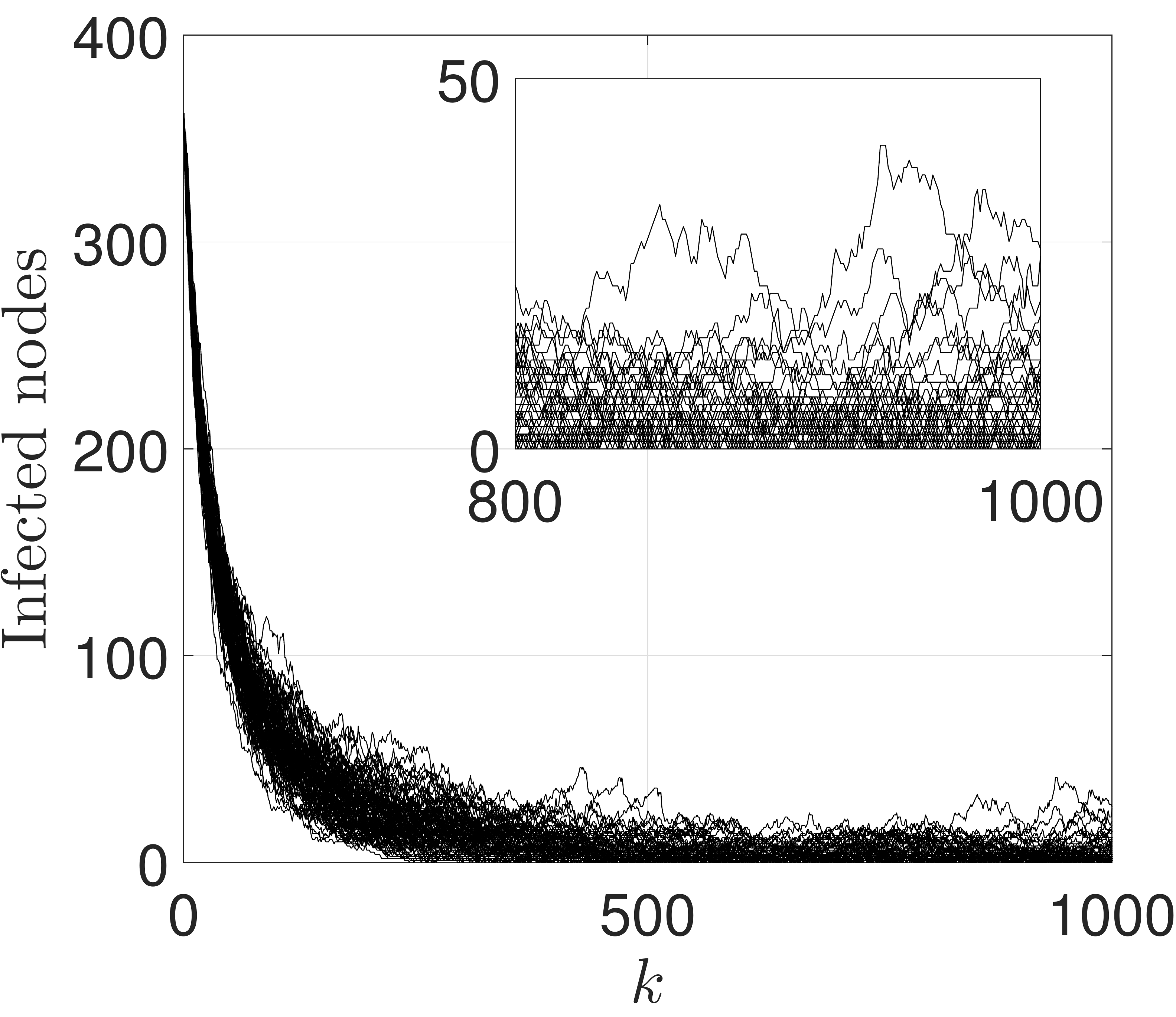}}
\caption{Sample paths of the number of infected nodes for different values of $\beta$.}
\label{fig:samplepath}
\end{figure*}

In this section, we illustrate the effectiveness of our results with numerical simulations. In our illustrations, we consider a discrete-time edge-independent dynamic random graph $\mathcal G$ with $n$ nodes. This random graph process is specified by the two following rules:
\begin{enumerate}
\item Construct an undirected Erd\H{o}s-R\'enyi random graph over the nodes of $\mathcal G$ \cite{ErdHos1959}. Denote the resulting $n\times n$ adjacency matrix by $A$.

\item If $A_{ij}=1$, then nodes $i$ and $j$ are connected in $\mathcal G$ via a stochastic link described by a $\{0,1\}$-valued time-homogeneous Markov chain $\sigma_{ij}$. The transition probability from $0$ to $1$ in this Markov process is given by $q_{ij}\geq 0$; while the transition probability from $1$ to $0$ is given by $r_{ij} \geq 0$. If $A_{ij}=0$, then nodes $i$ and $j$ are not connected in $\mathcal G$ at any time.
\end{enumerate}

One can easily deduce that, according to the second rule, $\bar A_{ij} = \bar A_{ji} = q_{ij}/(r_{ij} + q_{ij})$ if $A_{ij}=1$. In the following simulations, we let $n=500$ and consider a realization of an the Erd\H{o}s-R\'enyi graph with an edge probability of $0.2$ (in step 1 above).  The values of the transition probabilities~$r_{ij}$ (in step 2 above) are heterogeneous over the links of the Erd\H{o}s-R\'enyi graph. In particular, we have chosen for $r_{ij}$ a collection of random values from the Gaussian $\mathcal N(1/2,1/8)$, where we have truncated those values $r_{ij} >1$ to  $r_{ij} = 1$, and $r_{ij} < 0$ to $r_{ij} = 0$. Finally, we let $q_{ij} = 1 - r_{ij}$.

We consider a discrete-time spreading process with homogeneous infection and recovery rates, i.e.,  $\beta_i = \beta$ and $\delta_i = 0.05$ for all $i$. Using Theorem~\ref{thm:disc:edge}, we deduce that the disease-free equilibrium of $\Sigma_d$ is almost surely {exponentially} stable if $\beta < 6.32 \times 10^{-4}$. {The upper bound $\gamma_D$ of the decay rate versus $\beta$ is shown in Fig.~\ref{fig:upperBounds}.} In order to check the accuracy of this epidemic threshold, we compare this threshold with the one obtained from the following simulation. We simulate the spreading process over the dynamic random graph described above for various values of $\beta$. For each value of $\beta$, we generate $500$ sample paths of the spreading process. {In this simulation, we prevent a spreading process from dying out by re-infecting a randomly chosen node immediately after the infection process dies (i.e., when all the nodes become susceptible). A similar re-infection mechanism is employed for simulating spreading processes over static networks~\cite{Cator2013}. After the simulation, we compute the averaged number of infected nodes $y^*$ at time $k = 1,000$. This is the meta-stable number of infected nodes when re-infection of nodes is allowed. We then determine the metastable number of infected nodes in the original spreading process without reinfection as $z^* = y^*-1$, where the subtraction of one compensates the effect of re-infection. Finally, we define the empirical epidemic threshold $\beta^*$ as the maximum value of $\beta$ such that $z^* < 1$.}

Fig.~\ref{fig:beta-vs-fraction} shows the value of $z^*$ as $\beta$ varies. We see that the above defined empirical threshold~$\beta^*$ lies between $7.2\times 10^{-4}$ and $7.5\times 10^{-4}$, confirming that our condition indeed gives a sufficient condition for stability. On the other hand, the epidemic threshold based on the aggregated static network~$\bar A$ (i.e., the supremum of $\beta$ such that $\beta/\delta < 1/\eta(\bar A)$) is $9.95 \times 10^{-4}$ and, therefore, overestimates the actual epidemic threshold. In Fig.~\ref{fig:samplepath}, we show multiple realizations of the sample paths used to obtain Fig.~\ref{fig:beta-vs-fraction} for $\beta = 6.0 \times 10^{-4}$, $7.5 \times 10^{-4}$, and $9.0 \times 10^{-4}$.

\section{Proofs of Main Results}\label{sec:pf}

This section presents the proofs of the theorems presented in Section~\ref{sec:analysis}.

\subsection{Proof of Theorem~\ref{thm:links}} \label{app:arc}

We begin by recalling
a basic result on the stability analysis of a class of stochastic differential
equations called switched linear systems. Let $\sigma = f(\theta)$ be an
aggregated Markov process defined by the mapping $f \colon \Lambda \to \Gamma$
and a Markov process~$\theta$ with state space $\Lambda$. For each $\gamma \in
\Gamma$, there is an associated state matrix $A_\gamma \in \mathbb{R}^{n\times
n}$. An \emph{aggregated Markov jump linear system} is defined by the following
stochastic differential equation
\begin{equation}\label{eq:AMJLS}
\dot x(t) = A_{f(\theta(t))}x(t), 
\end{equation}
where $x(0) = x_0 \in \mathbb{R}^n$ and $\theta(0) = \theta_0 \in \Lambda$. We remark that, if $f$ is the identity mapping, then the system in \eqref{eq:AMJLS} is a Markov jump linear system~\cite{Costa2013}. We say that the system in~\eqref{eq:AMJLS} is \emph{almost surely {exponentially} stable} if {there exists $\lambda > 0$ such that} $\Pr(\limsup_{t\to\infty} t^{-1}\log\norm{x(t)} \leq -\lambda) = 1$ for all $x_0$ and $\theta_0$. {The supremum of $\lambda$ satisfying the above condition is called the \emph{decay rate}.} In order to prove Theorem~\ref{thm:links}, we will need the following criterion for almost sure {exponential} stability of aggregated Markov jump linear systems.

\begin{lemma}\label{lem:ascondition}
{Assume that $\theta$ is an irreducible Markov process and let $\pi$ be its
unique stationary distribution. Assume that
\begin{equation}\label{eq:as-cond}
E\left[\mu(A_{f(\pi)})\right] < 0. 
\end{equation}
Then, the aggregated Markov jump linear system
in \eqref{eq:AMJLS} is almost surely {exponentially} stable with a decay rate greater than or equal to $-E[\mu(A_{f(\pi)})]$}.
\end{lemma}

\begin{IEEEproof}
The above statement is known to be true for Markov jump linear systems \cite[Theorem~4.2]{Fang2002c}, i.e., when $f$ is the identity mapping. Let us now consider an arbitrary function $f$. Define $B_\lambda = A_{f(\lambda)}$ for each $\lambda \in \Lambda$. Then, we see that the system in \eqref{eq:AMJLS} is equivalent to the Markov jump linear system $\dot x(t) = B_{\theta(t)}x(t)$, for which we can guarantee almost sure {exponential} stability {with decay rate grater than or equal to $- E[\mu(B_\pi)]>0$} via the condition $E[\mu(B_\pi)] < 0$ \cite[Theorem~4.2]{Fang2002c}. This condition is equivalent to \eqref{eq:as-cond} by the definition of $B_\lambda$.
\end{IEEEproof}

Using Lemma~\ref{lem:ascondition}, we can prove the following preliminary
result, which will be useful in the proof of Theorem~\ref{thm:links}.

\begin{proposition}\label{prop:E[mu(M)]<0}
Consider an irreducible AMAI random graph process $\mathcal G$. For all distinct
and ordered pairs $(i,j)\in [n]^2$, let $h_{ij}$ be an independent Bernoulli
random variable with mean $\bar A_{ij}$. Define the random matrix
\begin{equation}\label{eq:M1}
M_1 = -D + \sum_{i=1}^n \sum_{j\neq i}\beta_i h_{ij} E_{ij}, 
\end{equation}
where $E_{ij}$ denotes the $\{0, 1\}$-matrix whose entries are all zero except
its $(i,j)$-entry. {If $E[\mu(M_1)] < 0$, then, the infection-free equilibrium of $\Sigma$ is almost
surely exponentially stable with decay rate greater than or equal to~$-E[\mu(M_1)]$.}
\end{proposition}

\begin{IEEEproof}
Define the direct product $\theta = \bigotimes_{i=1}^n \bigotimes_{j\neq
i}\theta_{ij}$, which is a stochastic process with state space $\Lambda =
\bigotimes_{i=1}^n \bigotimes_{j\neq i} \Lambda_{ij}$. Define \mbox{$f\colon
\Lambda \to \{0, 1\}^{n(n-1)}$} by $(f(\lambda))_{ij} = f_{ij}(\lambda_{ij})$.
Also, for $\gamma = (\gamma_{ij})_{i, j} \in \{0, 1\}^{n(n-1)}$, define the
matrix
\begin{equation}\label{eq:F_gamma}
F_\gamma = \sum_{i=1}^n \sum_{j\neq i} \gamma_{ij}E_{ij}. 
\end{equation}
Then, from Definition \ref{def:am:arc}, we see that $A(t) =
\sum_{i=1}^n\sum_{j\neq i} \sigma_{ij}(t)E_{ij} = F_{\sigma(t)} =
F_{f(\theta(t))}$. Therefore, we can rewrite $\bar{\Sigma}$ as the aggregated
Markov jump linear system $\Sigma_1 : \dot p(t) = (B F_{f(\theta(t))} - D)p(t)$.

By the irreducibility of $\mathcal G$, each time-homogeneous Markov process
$\theta_{ij}$ has a unique stationary distribution $\pi_{ij}$ on $\Lambda_{ij}$.
Then, the unique stationary distribution of~$\theta$ is \mbox{$\pi =
\bigotimes_{i=1}^n \bigotimes_{j\neq i} \pi_{ij}$}. Therefore, by
Lemma~\ref{lem:ascondition}, if $E[\mu(BF_{f(\pi)} - D)] < 0$, then $ \Sigma_1 =
\Sigma$ is almost surely {exponentially} stable {with decay rate greater than or equal to $-E[\mu(BF_{f(\pi)} - D)]$}. Therefore, to complete the proof of the
proposition, we need to show that
\begin{equation}\label{eq:...=M_1}
BF_{f(\pi)} - D = M_1, 
\end{equation}
in the sense that the random matrices appearing in the both sides of the
equation share the same probability distribution. From \eqref{eq:F_gamma} and
the definition of the matrix~$B$, we have
\begin{equation}\label{eq:F_f(pi)=...}
BF_{f(\pi)} -D = -D + \sum_{i=1}^n \sum_{j\neq i} \beta_i f_{ij}(\pi_{ij}) E_{ij}. 
\end{equation}
Since $f_{ij}$ maps into $\{0, 1\}$, the random variable $f_{ij}(\pi_{ij})$
is the Bernoulli random variable with mean equal to
\begin{equation*}
\begin{aligned}
\Pr\left(f_{ij}(\pi_{ij}) = 1\right) 
&= 
\pi_{ij}\left(f_{ij}^{-1}(\{1\})\right)
\\
&=
\lim_{t\to\infty} \Pr\left(\theta_{ij}(t) \in f_{ij}^{-1}(\{1\})\right)
\\
&=
\lim_{t\to\infty} \Pr( \sigma_{ij}(t) = 1)
\\
&=
\bar A_{ij}, 
\end{aligned}
\end{equation*}
where we used Lemma~\ref{lem:markov}. Moreover, all the random variables
$f_{ij}(\pi_{ij})$ ($i, j\in [n]$, $i\neq j $) are independent of each other
because $\mathcal G$ is arc-independent. From these observations and the
equation \eqref{eq:F_f(pi)=...}, we obtain \eqref{eq:...=M_1}, as desired.
\end{IEEEproof}

In order to complete the proof of Theorem~\ref{thm:links}, we need the following
result  from the theory of random graphs concerning the maximum eigenvalue of
the sum of random symmetric matrices.

\begin{proposition}[\cite{Chung2011}]\label{prop:Chung}
Let $X_1$, $\dotsc$, $X_N$ be independent random $n\times n$ symmetric matrices.
Let $C$ be a nonnegative constant such that $\norm{X_k-E[X_k]} \leq C$ for every
$k\in [N]$ with probability one. Also let $v^2 = \norm{\sum_{k=1}^N \Var(X_k)}$.
Then the sum $X = \sum_{k = 1}^N X_k$ satisfies
\begin{equation}\label{eq:Chung}
\Pr\left(\eta(X) >\eta(E[X]) + s\right) \leq \kappa_{C, v^2}(s)
\end{equation}
for every $s \geq 0$ (where $\kappa_{b,d}(s)$ was defined in \eqref{kappa_function}). 
\end{proposition}

\begin{IEEEproof}
Let $s \geq 0$ be arbitrary. Under the assumptions stated
in the proposition, it is shown in \cite{Chung2011} that, for every
$\theta>0$,
\begin{equation*}
\Pr\left(\eta(X) > \eta(E[X]) + s\right) 
\leq 
n\exp\left( -\theta s + \frac{1}{2}g(C \theta) \theta^2 v^2 \right), 
\end{equation*}
where $g$ is defined by $g(x) = 2x^{-2}(e^x-x-1)$ for $x>0$. It is easy to show
that the right hand side of the inequality takes its minimum value with respect
to $\theta$ when $\theta = (1/C)\log(1+(Cs/v^2))$. Substituting this particular
value of $\theta$ into the inequality, we readily obtain \eqref{eq:Chung}.
\end{IEEEproof}

In order to complete the proof of Theorem~\ref{thm:links}, we also need the next lemma concerning Metzler matrices. 

\begin{lemma}[{\cite[Lemma~2]{Son1996}}]\label{lem:MetMono}
Let $A$ and $B$ be Metzler matrices. If $A\leq B$, then we have $\eta(A) \leq
\eta(B)$ and $\mu(A) \leq \mu(B)$.
\end{lemma}


We can now provide a proof of Theorem~\ref{thm:links}.

\begin{IEEEproof}[Proof of Theorem~\ref{thm:links}]
Let $\lambda_1 = \mu(B\bar A - D)$ and define $X = M_1 + M_1^\top -
2\lambda_1 I - \kappa_{\bar \beta, \Delta_1}^{-1}\!(1) I$. Notice that 
\begin{equation}\label{eq:lammax0}
\eta(E[X]) = - \kappa_{\bar \beta, \Delta_1}^{-1}\!(1), 
\end{equation}
because $\eta(E[M_1 + M_1^\top]) = 2\mu(E[M_1]) = 2\mu(B\bar A - D) = 2\lambda_1$
by \eqref{eq:...=M_1}. Also, from the definition of $M_1$ in  \eqref{eq:M1}, we have
\begin{equation*}
X = \left(-2\lambda_1 I -2D - \kappa_{\bar \beta, \Delta_1}^{-1}\!(1)I\right) + \sum_{i=1}^n \sum_{j\neq i} X_{ij}, 
\end{equation*}
where $X_{ij} = \beta_i(E_{ij}+E_{ji})h_{ij}$ ($i, j\in [n]$ and $i\neq j$) are
random, independent symmetric matrices. We then apply
Proposition~\ref{prop:Chung} to the random matrix $X$, for which we  choose $C =
\bar \beta$. Clearly, the first constant term of $X$ has zero variance, while a
simple computation shows that \mbox{$\Var(X_{ij}) = \beta_i^2 \bar A_{ij}(1-\bar
A_{ij})(E_{ii} + E_{jj})$}. Therefore
\begin{equation}\label{eq:v^2}
\begin{aligned}
v^2 
&=
\biggl\lVert \sum_{i=1}^n \sum_{j\neq i} \beta_i^2 \bar A_{ij}(1-\bar A_{ij})(E_{ii} + E_{jj})\biggr\rVert
\\
&=
\biggl\lVert \bigoplus_{i=1}^n \biggl(\sum_{j=1}^n
\beta_i^2\bar A_{ij}(1-\bar A_{ij}) + \beta_j^2 \bar A_{ji}(1-\bar A_{ji})
\biggr)\biggr\rVert
\\
&= \Delta_1. 
\end{aligned}
\end{equation}
Combining \eqref{eq:Chung}, \eqref{eq:lammax0}, and \eqref{eq:v^2}, we
obtain the estimate
\begin{equation}\label{eq:Chung-special}
\Pr\bigl( \eta(X) > - \kappa_{\bar \beta, \Delta_1}^{-1}\!(1) + s\bigr) 
\leq 
\kappa_{\bar \beta, \Delta_1}(s)
\end{equation}
for $s > \kappa_{\bar \beta, \Delta_1}^{-1}\!(1)$. 

Let $(\Omega, \mathcal F, \Pr)$ be the fundamental probability space and
consider the set $\Omega_s = \{\omega \in \Omega \colon \eta(X) > - \kappa_{\bar
\beta, \Delta_1}^{-1}\!(1) + s \}$ for $s > \kappa_{\bar \beta,
\Delta_1}^{-1}\!(1)$. If $\omega \notin \Omega_s$, then $\eta(X) \leq -
\kappa_{\bar \beta, \Delta_1}^{-1}\!(1)+ s$ vacuously. On the other hand, if
$\omega \in \Omega_s$, then Lemma~\ref{lem:MetMono} gives the obvious estimate
\begin{equation*}
\begin{aligned}
\eta(X) 
&\leq 
2 \mu(B(\sgn \bar A) - D) - 2\lambda_1 -\kappa_{\bar \beta, \Delta_1}^{-1}\!(1)
\\
&= 
2c_1 - 2\lambda_1
\end{aligned}
\end{equation*}
because $M_1 \leq -D + \sum_{i=1}^n \sum_{j\neq i} \beta_i E_{ij} = B(\sgn \bar
A) - D$. Therefore, from \eqref{eq:Chung-special} it follows that
\begin{equation}\label{eq:derived}
\begin{aligned}
&E[\eta(X)] 
\\
\leq&
(- \kappa_{\bar \beta, \Delta_1}^{-1}\!(1) + s) \Pr(\Omega\backslash\Omega_s) + (2c_1 - 2\lambda_1) \Pr(\Omega_s)
\\
\leq&
- \kappa_{\bar \beta, \Delta_1}^{-1}\!(1) + s  + (2c_1 - 2\lambda_1) \kappa_{\bar \beta, \Delta_1}(s). 
\end{aligned}
\end{equation}
By the definition of the random matrix $X$, this inequality implies that
\begin{equation}\label{eq:almosthere}
2E[\mu(M_1)] \leq 2\lambda_1(1-\kappa_{\bar \beta, \Delta_1}(s)) + s + 2c_1 \kappa_{\bar \beta, \Delta_1}(s).
\end{equation}
 
Now we assume that condition \eqref{eq:threshold:arc} in the theorem holds.
Then, there exists {$s^* > \kappa_{\bar \beta, 
\Delta_1}^{-1}\!(1)$ such that
$2\left(1-\kappa_{\bar \beta, \Delta_1}(s^*)\right)\lambda_1 < -s^* -
2c_1\kappa_{\bar \beta, \Delta_1}(s^*)$. Therefore, when $s=s^*$, the inequality
\eqref{eq:almosthere} yields that
$E[\mu(M_1)] \leq \lambda_1(1-\kappa_{\bar \beta, \Delta_1}(s^*)) + (s^*/2) + 
c_1 
\kappa_{\bar \beta, \Delta_1}(s^*) < 0$
and hence we can complete the proof of
Theorem~\ref{thm:links} by Proposition~\ref{prop:E[mu(M)]<0}.}
\end{IEEEproof}

\begin{remark}
The upper bound $\bar s_1$ of the interval in the maximization problem
\eqref{eq:tao_c} does not play any role in the proof of Theorem~\ref{thm:links}.
In fact, the theorem holds true even if we replace the interval with the
infinite interval~$( \kappa_{\bar \beta, \Delta_1}^{-1}\!(1), \infty)$. The
reason why we introduce the upper bound is that 1) it makes it easy for us to
find the maximum and 2) it does not introduce conservativeness. To prove the
second statement, it is sufficient to show that the objective function in
\eqref{eq:tao_c} is less than $\mu(B \bar A - D)$ if $s>\kappa_{\bar \beta,
\Delta_1}^{-1}\!(1)$ and $s > \bar s_1$. From the former inequality, we obtain
$0< \kappa(s) <1$. Also, notice the that from Lemma~\ref{lem:MetMono} there
follows the inequality $-\ubar{\delta} = \mu(-D) \leq \mu(B\bar A -D)$.
Therefore, we can estimate the objective function as
\begin{equation*}
\begin{multlined}[.85\linewidth]
-\frac{s+2c_1\kappa_{\bar \beta, \Delta_1}(s)}{2(1-\kappa_{\bar \beta, \Delta_1}(s))}
<
-\frac{s}{2} - c_1\kappa_{\bar \beta, \Delta_1}(s)
\leq
\\
-\frac{\bar s_1}{2} + c_1^-
=
-\ubar{\delta}
\leq
\mu(B \bar A-D), 
\end{multlined}
\end{equation*}
as desired. We can justify the upper bounds $\bar s_2$ and $\bar s_3$ in the
maximization problems \eqref{eq:tao_E} and \eqref{eq:tao_H} in the same way.
\end{remark}

\begin{remark}
It can be easily verified that the proof of Theorem~\ref{thm:links} still holds
true even if we define $c_1$ to be $\mu(B(\sgn \bar A)- D)$. However, the
inclusion of the term~$-\kappa_{\bar \beta, \Delta_1}^{-1}\!(1)/2$ into $c_1$,
as was done in the theorem, makes the maximum in condition
\eqref{eq:threshold:arc} larger, and hence can reduce the conservativeness of
this condition.
\end{remark}

\subsection{Proof of Theorems~\ref{thm:edges} and \ref{thm:edge:homo}}\label{app:edge}

In this subsection, we prove Theorems~\ref{thm:edges} and
\ref{thm:edge:homo} for the edge-independent case in this subsection. We begin
with the following analogue of Proposition~\ref{prop:E[mu(M)]<0}.

\begin{proposition}\label{prop:E[mu(M)]<0:edge}
Consider an irreducible AMEI random graph process $\mathcal G$. Let $h_{ij}$
($1\leq i < j \leq n$) be independent Bernoulli random variables with mean $\bar
A_{ij}$. Define the random matrix
\begin{equation*}
M_2 = -D + \sum_{i=1}^n\sum_{j>i} \sqrt{\beta_i\beta_j} (E_{ij} + E_{ji}) h_{ij}. 
\end{equation*}
If
\begin{equation}\label{eq:suff-cond:edge}
E[\eta(M_2)] < 0, 
\end{equation}
then, the disease-free equilibrium of $\Sigma$ is almost surely 
{exponentially} stable with decay rate greater than or equal to $-E[\eta(M_2)]$.
\end{proposition}

\begin{IEEEproof}
Define the direct product $\theta = \bigotimes_{i=1}^n \bigotimes_{j>i}
\theta_{ij}$, which is a Markov process having the state space $\Lambda =
\bigotimes_{i=1}^n \bigotimes_{j>i} \Lambda_{ij}$. Also, define $g\colon \Lambda
\to \{0, 1\}^{n(n-1)/2}$ by $(g(\lambda))_{ij} = g_{ij}(\lambda_{ij})$. For
$\gamma = (\gamma_{ij})_{i,j} \in \{0, 1\}^{n(n-1)/2}$, define the matrix
\begin{equation} \label{eq:G_gamma}
G_\gamma = \sum_{i=1}^n\sum_{j>i} \gamma_{ij}(E_{ij} + E_{ji}). 
\end{equation}
Then, we can show that $A = \sum_{i=1}^n\sum_{j> i} \sigma_{ij}(E_{ij}+E_{ji}) =
G_{\sigma} = G_{g(\theta)}$ by Definition \ref{def:am:edge}. Moreover,
$g(\theta)$ is an aggregated Markov process, since $\theta$ is a Markov process.
Hence, we can rewrite $\Sigma$ as the aggregated Markov jump linear system 
$\dot p(t) = (B G_{g(\theta(t))} - D)p(t)$. The state transformation $x \mapsto
B^{-1/2} x$, where $B^{-1/2} = \diag (\beta_1^{-1/2}, \dotsc, \beta_n^{-1/2})$,
shows that almost sure {exponential} stability of this system is equivalent to almost sure {exponential}
stability of the following aggregated Markov jump linear system:
\begin{equation*}
\Sigma_2: \dot p(t)
= 
(B^{1/2} G_{g(\theta(t))} B^{1/2}- D)p(t). 
\end{equation*}
To prove almost sure {exponential} stability of $\Sigma_2$, by Lemma~\ref{lem:ascondition}, it is sufficient to prove that $E[\eta(B^{1/2} G_{g(\pi)} B^{1/2}- D)] < 0$, where we have used the fact that $\eta(A) = \mu(A)$ for a symmetric matrix $A$. On the other hand, in the same way as we did in the proof of Proposition~\ref{prop:E[mu(M)]<0}, we can prove that random matrices $B^{1/2} G_{g(\pi)} B^{1/2}- D$ and $M_2$ have the same probability distribution. Therefore, condition \eqref{eq:suff-cond:edge} is sufficient to guarantee almost sure {exponential} stability of $\Sigma_2$ and hence of $\Sigma$. {Moreover, from the above argument, it is straightforward to see that the decay rate of the almost sure exponential stability is greater than or equal to $-E[\eta(M_2)]$. This completes the proof of the proposition.}
\end{IEEEproof}

Let us prove Theorem~\ref{thm:edges}.

\begin{IEEEproof}[Proof of Theorem~\ref{thm:edges}]
Let $\lambda_2 = \eta(B\bar A - D)$ and define 
\begin{equation*}
\begin{aligned}
X &= 
M_2 - \lambda_2 I -
\kappa_{\bar \beta, \Delta_2}^{-1}\!(1) I
\\
&=
\left(-\lambda_2 I -D - \kappa_{\bar \beta, \Delta_2}^{-1}\!(1) I\right) +  \sum_{i=1}^n\sum_{j>i} X_{ij}, 
\end{aligned}
\end{equation*}
where $X_{ij} = \sqrt{\beta_i\beta_j} (E_{ij} + E_{ji}) h_{ij}$. We now apply
Proposition~\ref{prop:Chung} to the random matrix $X$, for which we choose $C =
\bar \beta$. Also, in the same way as we derived \eqref{eq:v^2}, we can show
$v^2 = \Delta_2$. Since $\eta(E[X]) = -\kappa_{\bar \beta, \Delta_2}^{-1}\!(1)$,
from \eqref{eq:Chung} we obtain \mbox{$\Pr\bigl( \eta(X) > - \kappa_{\bar \beta,
\Delta_2}^{-1}\!(1) + s\bigr) \leq \kappa_{\bar \beta, \Delta_2}(s)$}. Then, in
the same way as we derived \eqref{eq:derived}, we can show $E[\eta(X)] \leq -
\kappa_{\bar \beta, \Delta_2}^{-1}\!(1) + s  + (c_2 - \lambda_2) \kappa_{\bar
\beta, \Delta_2}(s)$, which implies $E[\eta(M_2)] \leq \lambda_2(1-\kappa_{\bar
\beta, \Delta_2}(s)) + s + c_2\kappa_{\bar \beta, \Delta_2}(s)$. Using this
inequality and Proposition~\ref{prop:E[mu(M)]<0:edge}, in the same way as we did
in the proof of Theorem~\ref{thm:links}, we can show that
condition~\eqref{eq:threshold:edges} in the theorem is indeed sufficient for
almost sure {exponential} stability of the disease-free
equilibrium of $\Sigma$ with decay rate greater than or equal to \eqref{eq:decayRate:AMEI}.
\end{IEEEproof}

Finally, we give an outline of the proof of Theorem~\ref{thm:edge:homo}
below.

\begin{IEEEproof}[Proof of Theorem~\ref{thm:edge:homo}]
Consider an irreducible AMEI random graph process $\mathcal G$. Assume homogeneous spreading and recovery rates, $\beta_i = \beta>0$ and $\delta_i = \delta>0$ for all $i$. In the same way as we did in the proofs of Propositions~\ref{prop:E[mu(M)]<0} and \ref{prop:E[mu(M)]<0:edge}, we can show that if $\eta(E[M_3] ) < \delta/\beta$ for $M_3 = \sum_{i=1}^n\sum_{j>i}(E_{ij} + E_{ji}) h_{ij}$, then the disease-free equilibrium of $\Sigma$ is almost surely {exponentially} stable with decay rate greater than or equal to $(\delta/\beta) - \eta(E[M_3])$. Let $\lambda_3 = \eta(\bar A)$ and define $X = M_3 - \lambda_3 I - \kappa^{-1}_{1, \Delta_3}(1)$. Applying Proposition~\ref{prop:Chung} to the random matrix~$X$, we obtain
\begin{equation*}
E[\eta(X)] \leq - \kappa_{1, \Delta_3}^{-1}\!(1) +  s  + (c_3-\lambda_3) \kappa_{1, \Delta_3}(s).
\end{equation*}
Therefore, $E[\eta(M_3)] \leq \lambda_3 + s + (c_3-\lambda_3)\kappa_{1, \Delta_3}(s)$. Using this inequality and the assumption~\eqref{eq:Cond}, we can actually show $\eta(E[M_3] ) < \delta/\beta$ and therefore the almost sure  {exponential} stability of $\Sigma$ in the same way as the proofs of Theorems~\ref{thm:links} and \ref{thm:edges}. We omit the details of the derivation of the lower bound~\eqref{eq:decayRateHomo} of the decay rate. Also, it is straightforward to show that \eqref{eq:beta/delta>eta(barA)} implies $\xi_H<1$. This completes the proof of the theorem.
\end{IEEEproof}

\subsection{Proof of Theorem~\ref{thm:disc:edge}} \label{app:disc}

In this subsection, we give the proof of
Theorem~\ref{thm:disc:edge} for spreading processes running in discrete-time.
In order to prove this theorem, we need to recall some facts regarding the
stability of switched linear systems. Let $\sigma = f(\theta)$ be an aggregated
Markov chain defined by the mapping $f \colon \Lambda \to \Gamma$ and a Markov
chain $\sigma$ with state space $\Lambda$. Let us define a state matrix
$A_\gamma \in \mathbb{R}^{n\times n}$ for each $\gamma \in \Gamma$. A
discrete-time aggregated-Markov jump linear system is described by the
stochastic difference equation
\begin{equation}\label{eq:AMJLS:disc}
x(k+1) = A_{f(\theta(k))}x, 
\end{equation}
where $x(0) = x_0$ and $\theta(0) = \theta_0$. We say that the system in~\eqref{eq:AMJLS:disc} is \emph{almost surely {exponentially} stable} if {there exists $\lambda > 0$ such that} \mbox{$\Pr(\limsup_{k\to\infty} k^{-1}\log\norm{x(k)} \leq -\lambda) = 1$} for all $x_0$ and $\theta_0$. {The supremum of $\lambda$ satisfying the above condition is called the \emph{decay rate}}. To prove Theorem~\ref{thm:disc:edge}, we need the following criterion for almost sure {exponential} stability of aggregated-Markov jump linear systems in discrete-time.

\begin{lemma}\label{lem:discstbl}
Consider an irreducible Markov chain $\theta$ and denote its unique stationary distribution by $\pi$. Assume that $A_\gamma$ is nonnegative and symmetric for every $\gamma \in \Gamma$. if $E\left[\log( \eta(A_{f(\pi)}))\right] < 0$, then the discrete-time aggregated Markov jump linear system in \eqref{eq:AMJLS:disc} is almost surely {exponentially} stable with decay rate greater than or equal to $-E[\log( \eta(A_{f(\pi)}))]$.
\end{lemma}

\begin{IEEEproof}
Let $\norm{A}$ denote the maximum singular value of a matrix $A$. Since $\eta(A) = \norm{A}$ for a nonnegative and symmetric matrix $A$, it is sufficient to show that \mbox{$E[\log\norm{A_{f(\pi)}}] < 0$} implies almost sure {exponential} stability of the aggregated Markov jump linear system~\eqref{eq:AMJLS:disc} with decay rate smaller than or equal to~$-E[\log\norm{A_{f(\pi)}}]$. This claim is known to be true if $f$ is the identity mapping, i.e., if $f(\theta)$ itself is a Markov chain \cite[Proposition~2.1]{Bolzern2004}. The proof for the general case where $f(\theta)$ is not necessarily a Markov chain has the same structure as the proof of Lemma~\ref{lem:ascondition}. We omit the details.
\end{IEEEproof}

Using this lemma, we prove the next proposition. 

\begin{proposition}\label{prop:}
Consider an irreducible and aperiodic AMEI random graph process $\mathcal G$ in
discrete-time. Let $h_{ij}$ ($1\leq i<j\leq n$) be independent Bernoulli random
variables with mean $\bar A_{ij}$. Define the random matrix
\begin{equation*}
M_4 = I-D + \sum_{i=1}^n\sum_{j>i} \sqrt{\beta_i\beta_j} (E_{ij} + E_{ji}) h_{ij}. 
\end{equation*}
If 
\begin{equation}\label{eq:ElognormM<0}
E\left[\log (\eta(M_4))\right] < 0, 
\end{equation}
then the disease-free equilibrium of $\Sigma_d$ is almost surely {exponentially} stable with decay rate greater than or equal to~$-E[\log (\eta(M_4))]$.
\end{proposition}

\begin{IEEEproof}
For each $k\geq 0$, let $A(k)$ denote the adjacency matrix of $\mathcal G(k)$. Define the product process $\theta$ on the state space $\Lambda$ in the same way as in the proof of Proposition~\ref{prop:E[mu(M)]<0:edge}. For $\gamma = (\gamma_{ij})_{i,j} \in \{0, 1\}^{n(n-1)/2}$, consider the matrix $G_\gamma$ defined in \eqref{eq:G_gamma}. Then, in the same way as in the proof of Proposition~\ref{prop:E[mu(M)]<0:edge}, we can show  $A(k) = G_{g(\theta(k))}$. Therefore $\Sigma$ is equivalent to the following discrete-time aggregated Markov jump linear system {$p(k+1) = (B G_{g(\theta(k))} + I - D) p(k)$}. Furthermore, the state transformation $x\mapsto B^{-1/2} x$ shows that almost sure {exponential} stability of this system is equivalent to that of the following aggregated-Markov jump linear system
\begin{equation*}
{\Sigma}_4 : p(k+1) = (B^{1/2}G_{g(\theta(k))}B^{1/2} + I - D) p(k). 
\end{equation*}
Since the matrix $B^{1/2}G_{g(\theta(k))}B^{1/2} + I - D$ is nonnegative and symmetric, by Lemma~\ref{lem:discstbl}, the system ${\Sigma}_4$ is almost surely stable if $E[\log (\eta(B^{1/2}G_{g(\pi)}B^{1/2} + I - D))] < 0$, where $\pi$ denotes the stationary distribution of the irreducible and aperiodic Markov chain $\theta$. On the other hand, in the same way as in the proof of Proposition~\ref{prop:E[mu(M)]<0:edge}, we can show that the random matrices $B^{1/2}G_{g(\pi)}B^{1/2} + I - D$ and $M_4$ have the same probability distribution. Therefore, condition \eqref{eq:ElognormM<0} is indeed sufficient for almost sure {exponential} stability of ${\Sigma}_4$, which implies almost sure {exponential} stability of the disease-free equilibrium of $\Sigma_d$ as we observed above. {Also, from the above discussion, it is easy to verify that $-E[\log (\eta(M_4))]$ gives a lower bound on the decay rate of stability.}
\end{IEEEproof}

We are now in condition to prove Theorem~\ref{thm:disc:edge}. 

\begin{IEEEproof}[Proof of Theorem~\ref{thm:disc:edge}]
Applying Proposition~\ref{prop:Chung} to the random matrix $M_4$, we obtain the
inequality $\Pr\left(\eta(M_4) > \eta(E[M_4]) + s \right) \leq \kappa_{\bar \beta,\Delta_2}(s)$.
Since $\lambda_4 = \eta(B\bar A + I - D) = \eta(E[M_4])$, this inequality implies that
\begin{equation*}
\Pr\left(\frac{\eta(M_4)}{\lambda_4} > 1 + \frac{s}{\lambda_4} \right)
\leq
 \kappa_{\bar \beta,\Delta_2}(s)
\end{equation*}
for every $s\geq 0$. Therefore, in the same way as we derived
\eqref{eq:derived}, we can show that
\begin{equation*}
E\left[\log \frac{\eta(M_4)}{\lambda_4}\right]
\leq
\log\left(
1 + \frac{s}{\lambda_4}
\right)
+ \kappa_{\bar \beta,\Delta_2}(s) \log \frac{\eta(M_{\max})}{\lambda_4}
\end{equation*}
and, hence, 
\begin{equation}\label{eq:ineq:disc}
E\left[\log(\eta(M_4))\right]
\leq
\log(
\lambda_4 + s
)
+ \kappa_{\bar \beta,\Delta_2}(s) \log \frac{\eta(M_{\max})}{\lambda_4}. 
\end{equation}
Therefore, if there exists $s \in [0, 1-\lambda_4]$ such that the right hand side of this inequality is negative, then disease-free equilibrium of $\Sigma_d$ is almost surely {exponentially} stable by Proposition~\ref{prop:}.  A simple algebra shows that the existence of such $s\geq 0$ is equivalent to \eqref{eq:disccond}. Notice that we do not need to consider any $s$ larger than $1-\lambda_4$ because, for such $s$, the right hand side of \eqref{eq:ineq:disc} is positive. Moreover, the upper bound~\eqref{eq:decayRateHomo} of the decay rate immediately follows from \eqref{eq:ineq:disc}. This completes the proof of the theorem.
\end{IEEEproof}

\section{Conclusion {and Discussion}} \label{sec:conclusion}

In this paper, we have analyzed the dynamics of spreading processes taking place over time-varying networks. First, we have proposed the family of aggregated-Markovian random graph processes as a flexible and analytically tractable dynamic random graph able to replicate, with arbitrary accuracy, any distribution of inter-switching times. We have then studied spreading processes in aggregated-Markovian random graph processes and derived conditions to guarantee that the disease-free equilibrium is almost surely {exponentially} stable. A direct analysis based on It\^o's formula for jump processes results in stability conditions in terms of the eigenvalues of a matrix whose size grows exponentially with the number of edges in the network. Using tools from random graph theory, we have derived alternative stability conditions in terms of the eigenvalues of a matrix whose size grows linearly with the number of nodes in the graph. Based on our theoretical results, we have shown that (\emph{i}) aggregated static networks approximate the epidemic threshold more accurately as the number of nodes in the network grows, and (\emph{ii}) aggregated static networks provide a better approximation as we reduce the degree of temporal variability in the random graph process.

{A possible direction for future research is containment of epidemic outbreaks over time-varying networks. Although several optimization frameworks have been recently proposed in the literature to find the cost-optimal allocation of medical resources to prevent epidemic outbreaks in static networks \cite{Wan2008IET,Preciado2013,Preciado2014,Drakopoulos2014}, these results cannot be readily applied to the case of time-varying networks.}





\ifCLASSOPTIONcompsoc
\section*{Acknowledgments}
\else
\section*{Acknowledgment}
\fi

This work was supported in part by the NSF under grants CNS-1302222 and IIS-1447470. A part of this research was performed while the first author was visiting the Center for BioCybernetics and Intelligent Systems at Texas Tech University. He would like to thank Professor Bijoy Ghosh
for his hospitality during this visit.

\ifCLASSOPTIONcaptionsoff
\newpage
\fi

\begin{IEEEbiography}[{\includegraphics[width=1in,height=1.25in,clip,keepaspectratio]{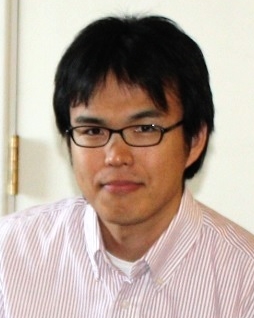}}]{Masaki Ogura}
received his B.Sc.~degree in engineering and M.Sc.~degree in informatics from Kyoto University, Japan, in 2007 and 2009, respectively, and his Ph.D.~degree in mathematics from Texas Tech University in 2014. He is currently a Postdoctoral Researcher
in the Department of Electrical and Systems Engineering at the University of Pennsylvania. His research interest includes dynamical systems on time-varying 
networks, switched linear systems, and stochastic processes.
\end{IEEEbiography}

\begin{IEEEbiography}[{\includegraphics[width=1in,height=1.25in,clip,keepaspectratio]{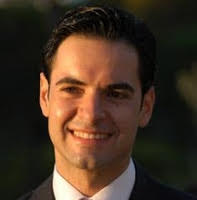}}]{Victor M.~Preciado} received his Ph.D. degree in Electrical Engineering and Computer Science from the Massachusetts Institute of Technology in 2008. He is currently the Raj and Neera Singh Assistant Professor of Electrical and Systems Engineering at the University of Pennsylvania. He is a member of the Networked and Social Systems Engineering (NETS) program and the Warren Center for Network and Data Sciences. His research interests include network science, dynamic systems, control theory, and convex optimization with applications in socio-technical systems, technological infrastructure, and biological networks.
\end{IEEEbiography}







\end{document}